\documentclass[aps,pre,onecolumn,showpacs,superscriptaddress,groupedaddress]{revtex4}
\usepackage{graphicx}
\usepackage{amsfonts}
\usepackage{bm}%
\usepackage[colorlinks=true,linkcolor=blue,filecolor=blue,menucolor=yellow,urlcolor=blue,
citecolor=blue,anchorcolor=blue]{hyperref}
\usepackage{graphicx}
\usepackage{dcolumn}
\usepackage{rotating}
\usepackage{subfig}
\usepackage[normalem]{ulem}
\usepackage{color}

\usepackage{amsmath}
\usepackage{ulem}
\usepackage{filecontents}
\usepackage[titletoc,toc,title]{appendix}
\usepackage[font=small,labelfont=bf,
   justification=justified,
   format=plain]{caption}

\usepackage[export]{adjustbox}
\usepackage{floatrow}

\begin{document}

\title{Tracer particles sense local stresses in an evolving multicellular spheroid without affecting the anomalous dynamics of the cancer cells}
\author{Himadri S. Samanta$^\oplus$}\affiliation{Department of Chemistry, University of Texas at Austin, TX 78712}
\author{Sumit Sinha$^\oplus$}\affiliation{Department of Physics, University of Texas at Austin, TX 78712}
\author{D. Thirumalai}\email{dave.thirumalai@gmail.com}\affiliation{Department of Chemistry, University of Texas at Austin, TX 78712}



\begin{abstract}
Measurements of local stresses on the cancer cells (CCs), inferred by embedding inert compressible tracer particles (TPs) in a growing multicellular spheroid (MCS), show that pressure decreases monotonically as the distance from the core of the MCS increases. How faithfully do the TPs report the local  stresses in the CCs is an important question because pressure buildup in the MCS is dynamically generated due to CC division, which implies that the CC dynamics should be minimally altered by the TPs. Here using theory and simulations, we show that  although the TP dynamics is unusual, exhibiting sub-diffusive behavior on times less than the CC division times and hyper-diffusive dynamics on in the long-time limit, they do not affect the long-time CC dynamics or the local CC stress distributions. The CC pressure profile within the MCS, which decays from a high value at the core to the periphery, is almost identical with and without the TPs.  That the TPs have insignificant effect on the local stresses in the MCS implies that they are  reliable reporters of the CC microenvironment.  
\end{abstract}

\date{\today}
\maketitle

\def\thefootnote{$\oplus$}\footnotetext{: Equal Contribution}\def\thefootnote{\arabic{footnote}}

The interplay between  short-range forces and non-equilibrium processes arising from cell division and apoptosis  gives rise to unexpected dynamics in the collective migration of cancer cells \cite{Friedl09NatRevMolCellBiol,Shaebani20NatRevPhys,kumar2009mechanics,desoize2000multicellular,Angelini4714, walenta2000metabolic, laurent2013multicellular, valencia2015collective,Han20NP}. An example is the dynamics of cancer cells (CCs) in a growing multicellular spheroid (MCS), which is relevant in cancer metastasis \cite{laurent2013multicellular, valencia2015collective}. Imaging experiments show that collective migration of a group of cells that maintain contact for a long period of time exhibits far from equilibrium characteristics \cite{valencia2015collective, richards20184d,martino2019wavelength, han2019cell, palamidessi2019unjamming}. 
The experimental studies have served as an impetus to develop simulations and theoretical models \cite{Abdul17Nature,Himadri18PRE,sinha2019spatially,Bi15NP,Bi16PRX,Park15NM}, which have given insights into the dynamics of collective motion in the MCS. Dynamics in a growing MCS is  reminiscent of the influence of active forces in abiotic systems \cite{marchetti2013hydrodynamics, bechinger2016active,Nandi18PNAS}. In a growing MCS, the analogue of active forces are self-generated \cite{sinha2020self}, arising from biological events characterized by cell growth, division and apoptosis.  

Although it has long been recognized that the stresses in the tumor interior are enhanced\cite{BoucherCR90,HelmlingerNB97,Jain14ARBE}, which explained the reduction in proliferation, it is only recently that direct non-uniform pressure variations in MCS has been measured. Following the early pioneering experiments \cite{BoucherCR90,HelmlingerNB97,Jain14ARBE}, several experimental studies that probe the local stresses or pressure on the CCs \cite{rauzi2011cortical,hutson2003forces,boucher1990interstitial,fadnes1977interstitial,campas2014quantifying,Dolega17NC} have provided insights into the mechanism by which the CCs invade the extracellular matrix. The stresses within MCSs were measured \cite{Dolega17NC} by embedding micron-sized inert deformable polyacrylamide beads, referred to as tracer particles (TPs). ~The reduction in the volume of the TPs in the presence of the colon carcinoma cancer cells, was used to estimate the strain on the TPs. From the calibrated stress-strain plot for the TPs, measured in the absence of the CCs, they obtained the stress value at the location of the TPs. By assuming that the stresses associated with the fluorescently labeled TPs faithfully report the pressure of the CCs, it was argued that the pressure propagates non-uniformly across the tumor. It was concluded that the TPs could be used as local stress probes or sensors. 
 
The experiments raise an important question:  \textit{What should be the characteristics of the TPs for them to be faithful sensors of stresses in an evolving MCS?} This question emerges naturally from the experimental results \cite{Dolega17NC,Mohagheghian18NC}. We believe that the following criteria must be satisfied : (a) Because in a growing MCS, pressure is dynamically generated predominantly by cell division, the TPs should have negligible effect on the CC dynamics. (b) The radial pressure profile in the MCS with and without the TPs must be similar. (c) Finally, the TPs should have negligible effect on the distribution of pressure on the cells across the MCS. In essence, the TPs should ``observe'' but not ``disturb'' the CC microenvironment.

 \floatsetup[figure]{style=plain,subcapbesideposition=top}
\begin{figure}
{\includegraphics[width=0.5\linewidth] {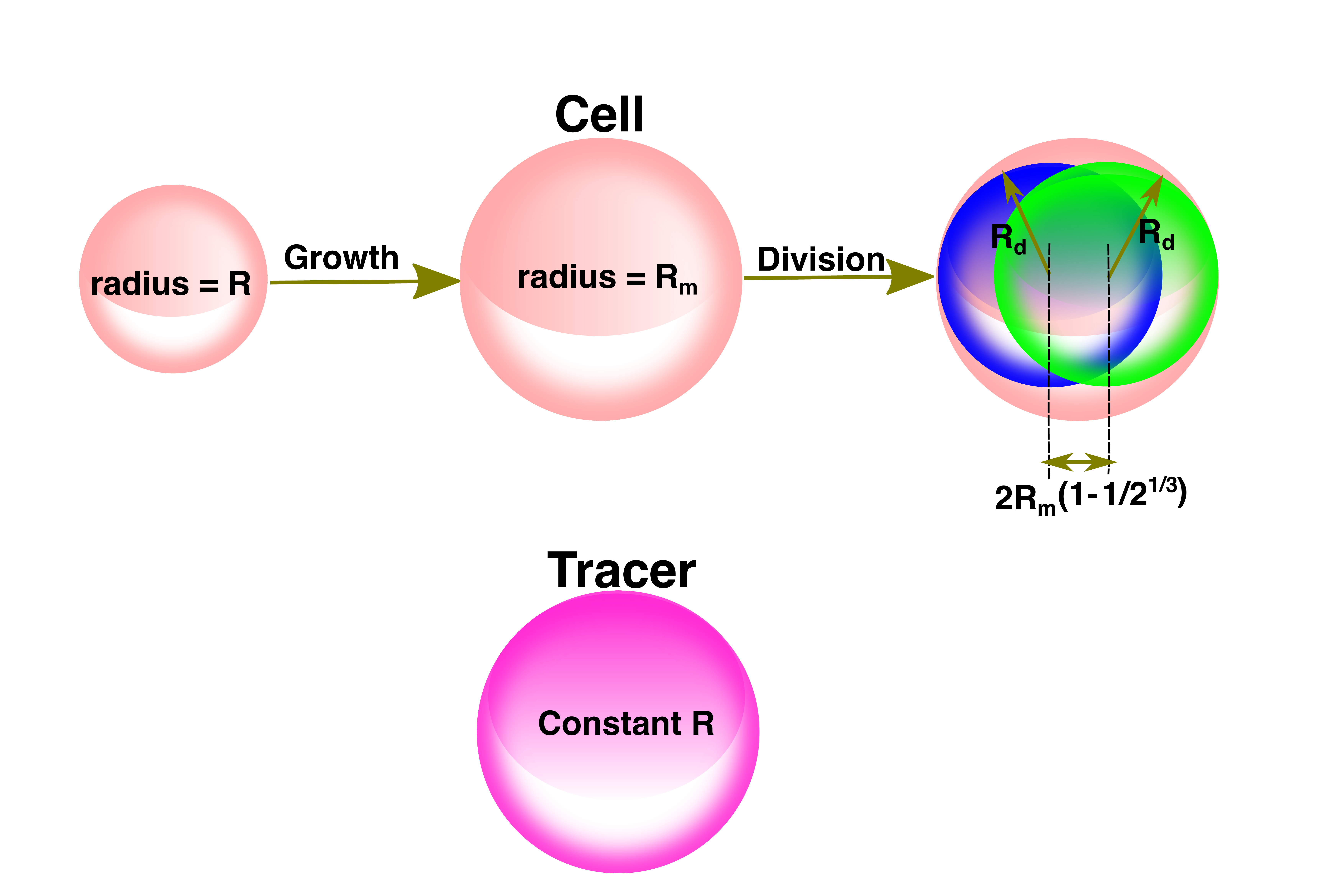}} 
\vspace{-.1 in}
\caption{Difference between the CCs (pink sphere) and the TPs (magenta sphere). Cell division creates two daughter CCs (blue and green sphere), each with $R_d=\frac{R_m}{2^{1/3}}$. Their relative position is displaced by $2R_m(1-\frac{1}{2^{1/3}})$ with the orientation being random with respect to the sphere center. }
\label{tracer_cell}
\end{figure}

In order to assess the influence of TPs on the fate of cancer cells, we used theory and simulations by embedding the TPs in an evolving MCS. We first show that the CCs, whose dynamics is sub-diffusive for $t \lesssim \tau$, exhibits super-diffusive behavior at long times with the MSD, $\Delta_{CC} (t)\sim t^{\alpha_{CC}}$. Surprisingly, $\alpha_{CC}$ does not change appreciably from the value in the absence of the TPs. 
Thus, criterion (a) described above is satisfied. We note parenthetically that the motility of the TPs is unusual. On short times ( $t \lesssim \tau$, the cell division time) 
$\Delta_{TP} \sim t^{\beta_{TP}}$ with $\beta_{TP} < 1$. In the long-time limit, ($t>\tau$), the TPs undergo hyper-diffusive motion, $\Delta_{TP}(t)\sim t^{\alpha_{TP}}$ with $\alpha_{TP}>2$.  We find that the pressure profiles vary radially across the MCS. The pressure is larger near the core and decreases as the boundary of the tumor is approached. Most importantly, the radial pressure profiles of the CCs with and without the TPs are virtually identical. We summarize that criteria (b) and (c) are satisfied, which shows that deformable gel-like TPs may be used as sensors of local stresses in the tumor.\\


\noindent{\bf Results:}\\
The stresses in an evolving tumor are generated by dividing cancer cells. The stresses, created by active forces generated CC division,  do not relax fast rapidly,  especially when the tumor acquires has core structure. Because the CCs generate active forces through stochastic cell division, we first begin by describing the TPs and CCs, and how they affect each other.\\

\begin{figure}
\subfloat[]{\includegraphics[width=0.55\linewidth] {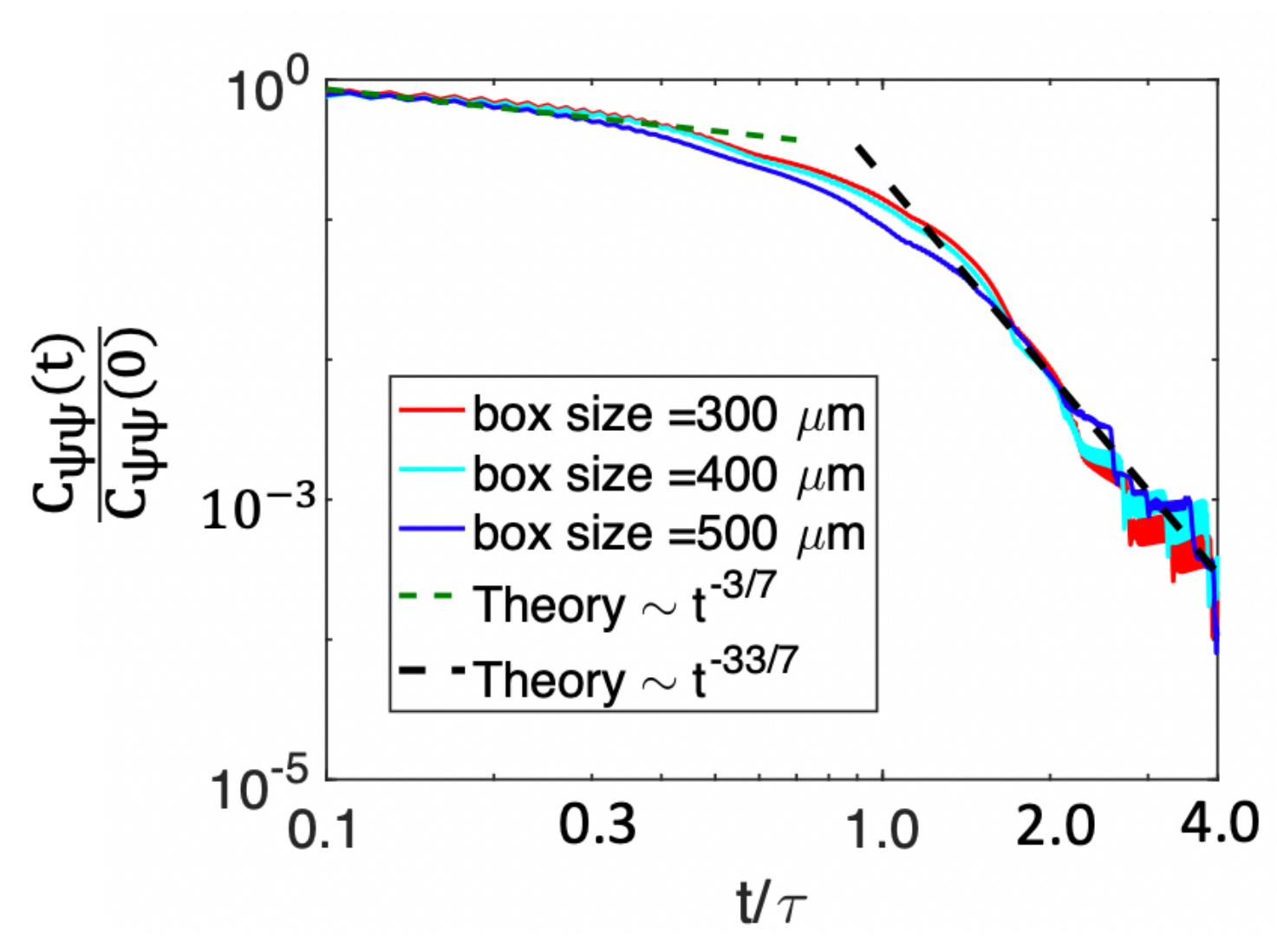}\label{fig1ac}}\\
\subfloat[]{\includegraphics[width=0.45\linewidth] {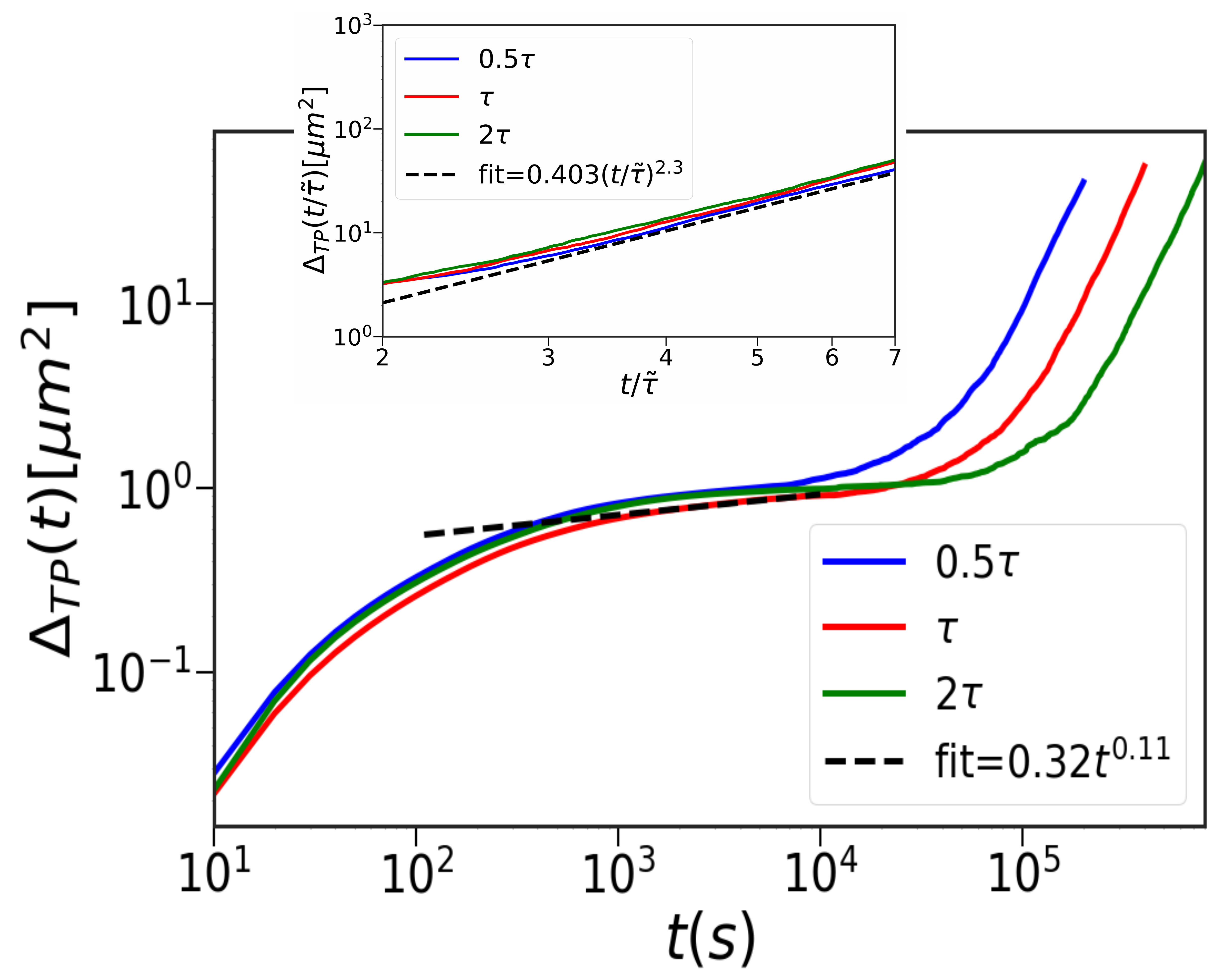}\label{fig1b}} 

\caption{(a) The normalized time-dependent density correlation function plot for the TPs. The cell cycle time is $\tau=54,000 ~\text{s}$. The integrated time ($t \approx 4\tau$) is long enough to obtain accurate exponents from the power law decay for both short ($t<\tau$) and long times ($t>\tau$). The plots for three different box sizes confirm that the values of the exponents do not depend on the system size. (b) MSD of the TPs ($\Delta_{TP}$) using the Gaussian potential. The curves are for  3 cell cycle times (blue ($0.5\tau$), red  ($\tau$), and   green ($2\tau$).  Time  to reach the {hyper-diffusive} behavior, {which is preceded by a jamming regime ($\Delta_{TP}\sim t^{\beta_{TP}}, \beta_{TP}$= 0.11, shown in black dashed line)}, increases with $\tau$.  The inset focusses on the hyper-diffusive regime ($\frac{t}{\tilde{\tau}}>1$). $\tilde{\tau}$ is the cell cycle time for the respective curves. Time is scaled by $\frac{1}{\tilde{\tau}}$. $\Delta_{TP} \sim t^{\alpha_{TP}}, \alpha_{TP}=2.3$ (dashed black line).}
\label{log_grow}
\end{figure}

\noindent{\it{\bf Theoretical Considerations}}:{\color{black}
~We consider the dynamics of deformable inert TPs in a growing MCS in a dissipative environment by neglecting inertial effects.
In contrast to the TPs, the CCs grow and divide at a given rate, and also undergo apoptosis (see Figure (\ref{tracer_cell})). The CCs and TPs experience systematic short-ranged attractive and repulsive forces arising from the other CCs and TPs \cite{Abdul17Nature}.  To model a growing MCS, we modify the density equation for the CCs phenomenologically by adding a non-linear source term, $\propto \phi(\phi_0-\phi)$, accounting for cell birth and apoptosis, and a non-equilibrium noise term that breaks the CC number conservation \cite{Gelimson15PRL}. 
The noise, $f_\phi$, satisfies $<f_\phi({\bf r},t)f_\phi({\bf r'},t')>=\delta({\bf r}-{\bf r}')\delta(t-t')$.
The source term, $\propto \phi(\phi_0-\phi)$, represents the birth and apoptosis, with  $\phi_0=\frac{2k_b}{k_a}$\cite{Doering03PA,Gelimson15PRL}.
The coefficient, $\sqrt{k_b \phi+\frac{k_a}{2}\phi^2} $ of $f_\phi$, is the noise strength, corresponding to number fluctuations. 

In order to obtain analytic insights, we first derive suitable equations that describe the dynamics of the TPs and CCs. We introduce the density fields $\phi({\bf r},t)=\sum_i \delta[{\bf r}-{\bf r}_i(t)]$ for the CCs, and $\psi({\bf r},t)=\sum_i \delta[{\bf r}-{\bf r}_i(t)]$ for the TPs.  A formally exact Langevin equation for $\phi({\bf r},t)$} and $\psi({\bf r},t)$ may be derived using the Dean's method~\cite{Dean96JPA} that accounts for diffusion and non-linear interactions. 
The equations for $\phi({\bf r},t)$ and $\psi({\bf r},t)$ are,
\begin{eqnarray}
\frac{\partial \psi({\bf r},t)}{\partial t}&=&D_\psi \nabla^2 \psi({\bf r},t)+ 
{\bf \nabla }\cdot \left(\psi({\bf r},t){\bf J}\right)+\tilde{\eta}_\psi , \label{trdensity}\\
\frac{\partial \phi({\bf r},t)}{\partial t}&=&D_\phi \nabla^2 \phi({\bf r},t)+ 
{\bf \nabla }\cdot \left(\phi({\bf r},t){\bf J}\right) +\frac{k_a}{2} \phi(\frac{2k_b}{k_a}-\phi)+\sqrt{k_b \phi+\frac{k_a}{2} \phi^2} f_\phi+\tilde{\eta}_\phi
 \,
\label{phi10}
\end{eqnarray}
where ${\bf J}=\int_{\bf r'}[\psi({\bf r'},t)+ \phi({\bf r'},t)]{\bf \nabla}U({\bf r-\bf{r'}})$, $\tilde{\eta}_\psi({\bf r},t)={\bf \nabla} \cdot \left(\eta_\psi({\bf r},t) \psi^{1/2}({\bf r},t)\right)$, $\tilde{\eta}_\phi({\bf r},t)={\bf \nabla} \cdot \left(\eta_\phi({\bf r},t) \phi^{1/2}({\bf r},t)\right)$, and $\eta_{\phi,\psi}$ satisfies $<\eta_{\phi,\psi}({\bf r},t)\eta_{\phi,\psi}({\bf r'},t')>=\delta({\bf r}-{\bf r}')\delta(t-t')$. 
{{The second term in Eq.(\ref{trdensity}) accounts for the TP-TP interactions (${\bf \nabla }\cdot \left(\psi({\bf r},t)\int_{\bf r'} \psi({\bf r'},t){\bf \nabla}U({\bf r-\bf{r'}})\right)$) and TP-CC interactions, (${\bf \nabla }\cdot \left(\psi({\bf r},t)\int_{\bf r'} \phi({\bf r'},t){\bf \nabla}U({\bf r-\bf{r'}})\right)$).} The influence of the CCs on the TP dynamics is reflected  in the TP-CC coupling. 
The third term in Eq.(\ref{phi10}) results from cell birth and apoptosis, and the fourth term in Eq.(\ref{phi10}) is the non-equilibrium noise.~It should be noted that Eq.(\ref{trdensity}) does not satisfy the  fluctuation-dissipation theorem in the tumor growth phase. 

The coupled stochastic integro-differential (SID) equations are difficult to solve analytically. Even numerical solutions in the absence of TPs, are obtained by replacing $\tilde{\eta}_\phi ({\bf r},t)$, the last term in Eq.(\ref{phi10}), by 
$\tilde{\eta}_\phi ({\bf r},t)=\nabla \cdot (\tilde{\eta}_\phi ({\bf r},t) \phi_0^{1/2})$, where $\phi_0$ is a constant.~Instead of using uncontrolled approximations, we solved the exact coupled SIDs (Eq.(\ref{trdensity}) and Eq.(\ref{phi10})) numerically (see the methods section for details) 
in order to calculate the needed correlation functions from which the exponents characterizing the mean square displacement may be derived. For instance, from the time-dependent decay of the correlation function 
$C_{\psi \psi}(t)= \int d^3r C_{\psi\psi}({\bf r},t)$, we can extract the dynamical exponent, $z$. Previously \cite{Abdul17Nature}, we had shown  that $C_{\psi\psi}(t)\sim t^{1-(2+d)/z}$ ($d$ is the spatial dimension).
The correlation function $C_{\psi\psi}(t)/C_{\psi\psi}(0)$ decays in two stages, depending on $t/\tau$, where $\tau$ is the cell division time (Figure(\ref{log_grow}a)). For $t/\tau<1$, we find that 
$C_{\psi\psi}(t)\sim t^{-3/7}$ from which we obtain $z=7/2$. The MSD ($\Delta_{TP}(t)$) exponent for the TPs is related to $z$ as $\Delta_{TP}(t)\sim t^{\beta_{TP}} \sim t^{2/z}$, which implies $\beta_{TP}=4/7$. 
Similarly, at long ($t/\tau>1$) times $C_{\psi\psi}(t) \sim t^{-33/7}$ from which we deduce that $\Delta_{TP}(t)\sim t^{\alpha_{TP}}$ with $\alpha_{TP}=16/7$. Thus, the numerical solution of the exact equations together with the scaling ansatz predict that, in the long time, the TPs undergo hyper-diffusive motion in the presence of CCs. }\\



\floatsetup[figure]{style=plain,subcapbesideposition=top}

 
\noindent{\it{\bf Simulations}}: In order to validate the theoretical predictions and decipher the mechanisms of the origin for CC driven hyper-diffusive motion, we simulated a 3D MCS with embedded TPs using an  agent-based model \cite{drasdo2005single, Abdul17Nature,sinha2019spatially,sinha2020self,sinha2021inter,sinha2021memory}. The cancer cells, treated as interacting soft deformable spheres, grow with time, and divide into two identical daughter cells upon reaching a critical mitotic radius ($R_m$). We used, $\tau = \frac{1}{k_b}=54,000s$ as the cell division time. The CCs also undergo apoptosis at the rate $k_a < < k_b$.  The sizes and the number of TPs are held constant. 

We used a pressure inhibition mechanism to model the observed growth dynamics in solid tumors \cite{Abdul17Nature}. Dormancy or the growth phase of the CCs  depends on the local microenvironment, which is determined by the pressure on the $i^{th}$ cell (Figure(\ref{tracer_cell})).  Cell division and the placement of the daughter cells stochasticity alter the dynamics of the CCs. More importantly, cell division generates active forces, referred to self-generated active force~\cite{sinha2020self}, that persist for a period of time, thus affecting the dynamics of the neighboring cells. To determine the TP dynamics, we model the CC-CC, CC-TP, and TP-TP interactions using two potentials (Gaussian and Hertz) to ensure that the qualitative results are robust (simulation details are in the methods section). 






\bigskip

\noindent{\it {\bf Mean Square Displacement of the TPs ($\Delta_{TP}(t)$):}} {{The simulations show that, in the limit $t<\frac{1}{k_b}$, the TPs exhibit sub-diffusive behavior with $\Delta_{TP}(t)\sim t^{\beta_{TP}}$ with $\beta_{TP}<1$ (Figure(\ref{log_grow}b)). Thus, on time scales less than $\tau$ the TPs exhibit sub-diffusive behavior, as predicted theoretically (\ref{log_grow}a)).
We also calculated $\Delta_{TP}(t)=\langle[{\bf r}(t)-{\bf r}(0)]^2\rangle$, by averaging over $\approx 2,000$ trajectories, for the Gaussian (Figure(\ref{log_grow}b)) and Hertz (Figure {\color{black}S3a in the SI}) potentials.  Similar behavior is found in both the cases. The modest increase in the MSD at short times is seen as a near plateau in $\Delta_{TP}(t)$ as a function of $t$ (Figure(\ref{log_grow}b)). The duration of the plateau ({Figure(\ref{log_grow}b)) increases as the cell cycle time increases. Although  jamming behavior predicted theoretically is consistent with the simulations the numerical values of $\beta_{TP}$ differ (compare Figure(\ref{log_grow}a) and (\ref{log_grow}b)). 
The dynamics in the $t/\tau<1$ limit depends on the details of the model.


\floatsetup[figure]{style=plain,subcapbesideposition=top}
\begin{figure}
\includegraphics[width=0.55\linewidth] {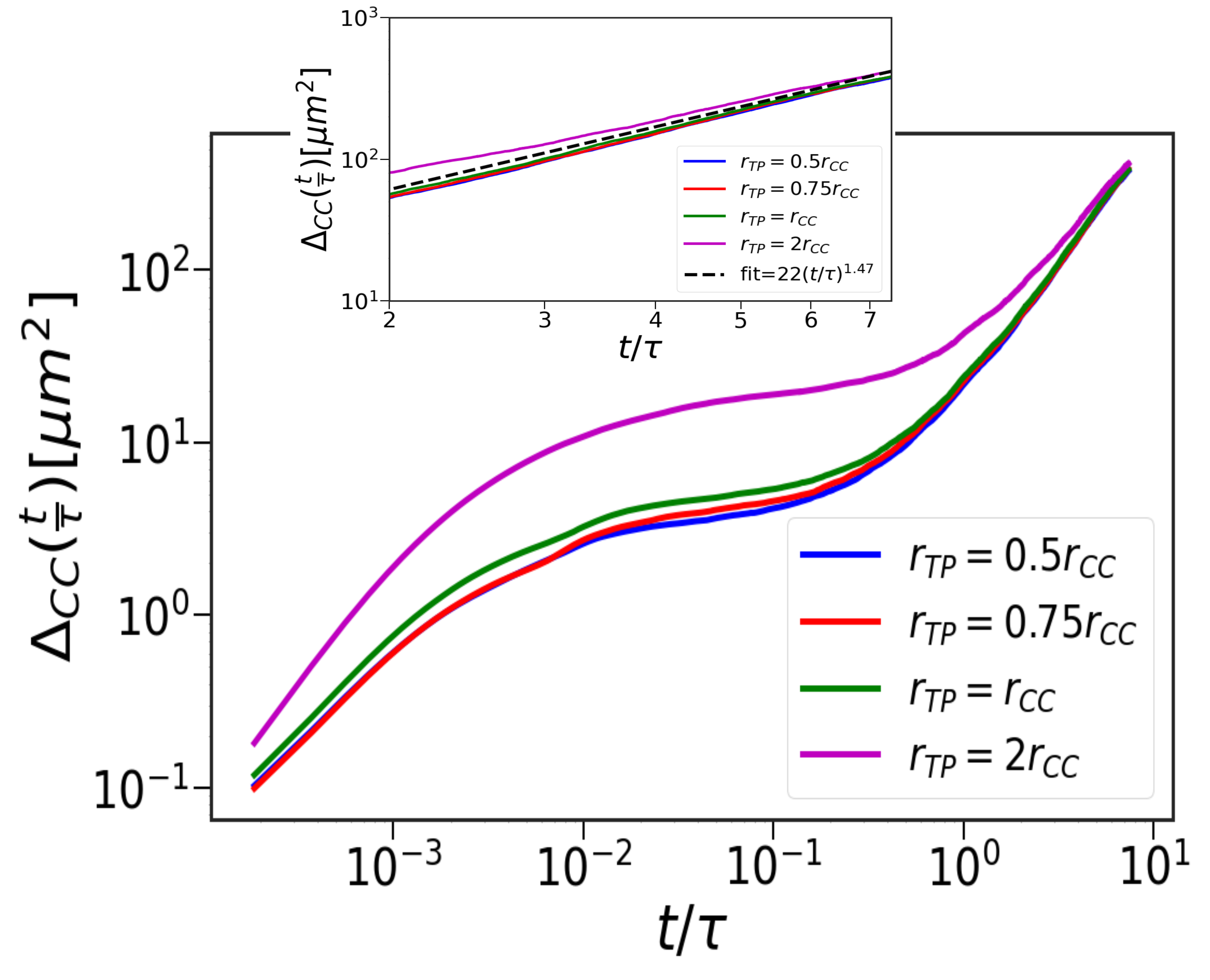}
\vspace{-.2 in}
\caption{{\color{black} CC dynamics in the presence of the TPs. $\Delta_{CC}$ using the Hertz potential. The curves are for different TP radius ( magenta - $r_{TP}=$ { 2} $r_{CC}$, green - $r_{TP} = r_{CC}$, and red - $r_{TP} = 0.75r_{CC}$ , and blue - $r_{TP} = 0.5r_{CC}$, where $r_{CC} =4.5 \mu m$ is the average cell radius. Inset shows  $\Delta_{CC}$ for the four curves, focusing on the super-diffusive regime.  The black dashed line is drawn with  $\alpha_{CC}=1.47$.}}
\label{fig4aa}
\end{figure}

In the $t>\frac{1}{k_b}$ limit, theory and simulations predict hyper-diffusive dynamics for the TPs, $\Delta_{TP} \sim t^{\alpha_{TP}}$ with  $\alpha_{TP}\approx 2.3$. 
{
Variations in $k_b=\tau^{-1}$ do not change the value of $\alpha_{TP}$ (Figure(\ref{log_grow}b)). It merely changes the coefficients of the linear term and the value of $\frac{2k_b}{k_a}$. Therefore, $\alpha_{TP}$  is independent of the cell cycle time in the long time limit.}
Simulation results are in excellent agreement with the theoretical predictions.  For the Gaussian potential, we obtain $\alpha_{TP} \approx$  2.3  (see the inset in Figure(\ref{log_grow}b)) for $0.5\tau$, $\tau$ and $2\tau$. Thus, $\alpha_{TP}$ is independent of cell cycle time, as argued above. 
{\color{black} Note that TP-TP interactions play an insignificant role in the dynamics of TPs or the CCs (Figure S4 in the SI). They merely alter the amplitude of $\Delta_{TP}$ in the intermediate time. They do not affect  the long time dynamics.
We conclude that the long-time behavior is universal, and is impervious to the details of the interaction but is determined by cell division and apoptosis times.}

\floatsetup[figure]{style=plain,subcapbesideposition=top}
\begin{figure}
{\includegraphics[width=0.5\linewidth] {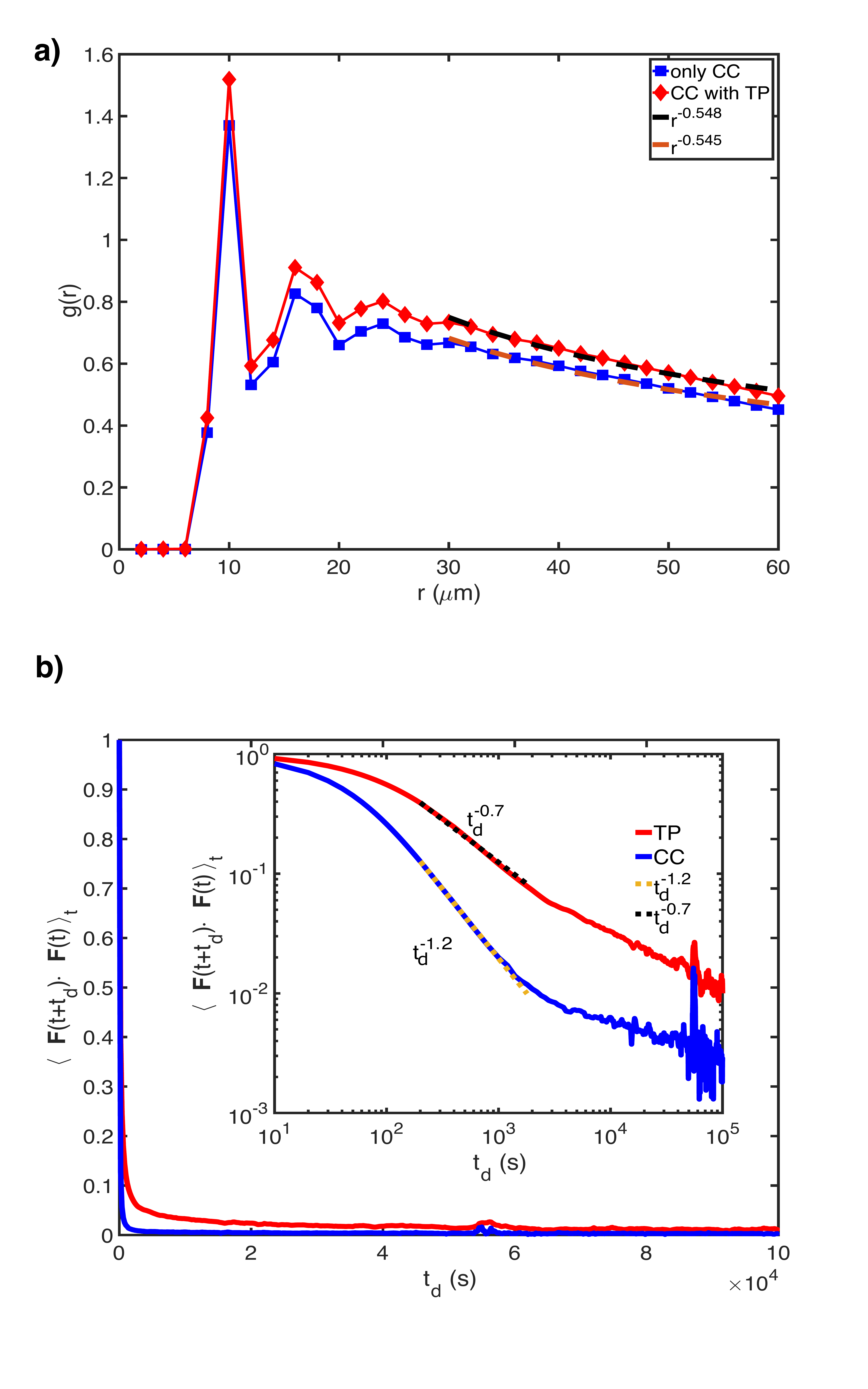}} 
\vspace{-.3 in}

\caption{ {\color{black} {\bf (a)} CC correlation function, ($g(r)$), as a function of inter-cellular distance. Red (blue) curve shows $g(r)$ in presence (absence) of TPs. The two dashed lines (black and orange) are power law fits to $g(r)$ in the large $r$ limit. {\bf (b)} Force force autocorrelation (FFA), as a function of delay time ($t_d$). The red (blue) curve shows the FFA for TPs (CCs). Inset shows FFA on log-log scale. The black (yellow) dashed line is a  power law fit with exponent of -0.7 (-1.2). } }
\label{gr_ffa}
\end{figure}

\bigskip
\noindent{\bf Impact of TP on CC dynamics:} Interestingly, the time-dependent changes in $\Delta_{CC}(t)$ are not significantly affected by the TPs. Figure(\ref{fig4aa}) shows that changing the tracer size affects only the amplitude of the $\Delta_{CC}$ in the intermediate time without altering the $\alpha_{CC}$ values (see also figure S5 in the SI). This is because the TP size only changes the nature of the short-range interactions without introducing any new length scale.  Since, $\alpha_{CC}$ is a consequence of the long-range spatial and temporal correlations that emerge because of the rates cell division and apoptosis, it is independent of the tracer size (see the details in the SI).

{\color{black}In the absence of the TPs,  birth and apoptosis determine the CC dynamics in the long time regime, yielding $\alpha_{CC}=1.33$ \cite{Abdul17Nature,Himadri18PRE}. 
When the TPs are present, the CCs continue to exhibit super-diffusive motion with a modest increase in $\alpha_{CC}$ (Table II in the SI).    The origin of the predicted super-diffusive behavior in the CCs is related to the long range spatial correlation that arises due the SGAF that is related to cell division. This is reflected in the CC pair-correlation, $g(r)=\frac{V}{4\pi r^2 N^2} \sum_{i=1}^{N}\sum_{j\neq i}^{N}\delta(r-|{\bf r}_i- {\bf r}_j|) \sim r^{-0.5}$, in the presence and absence of TPs at $t\approx 8 \tau$  (Figure(\ref{gr_ffa}a)).  The dynamically-induced long-range CC correlations is independent of the TPs, thus explaining the insignificant effect of the  TPs on  the CC dynamics. }
\bigskip

\noindent{\bf Mechanistic Origin:~}{\color{black} To explain the finding that, $\alpha_{TP}>\alpha_{CC}$, we calculated the force-force autocorrelation function, FFA=$\langle {\bf F}(t+t_d)\cdot {\bf F}(t)\rangle_t$. Here, $\langle ... \rangle_t$ is the time average and $t_d$ is the duration of the delay time. Since, the TPs (CCs) exhibit hyper-diffusion (super-diffusion), we expect that the FFA of TPs should decay slower relative to the CCs. This is confirmed in  Figure(\ref{gr_ffa}b), which  shows that the FFA for the TPs (CCs) decays as $t_d^{-0.7}$ ($t_d^{-1.2}$).  Thus, the TP motion is significantly more persistent than the CCs, which explains the hyper-diffusive nature of the TPs.}

\floatsetup[figure]{style=plain,subcapbesideposition=top}
\begin{figure}
{\includegraphics[width=0.71\linewidth] {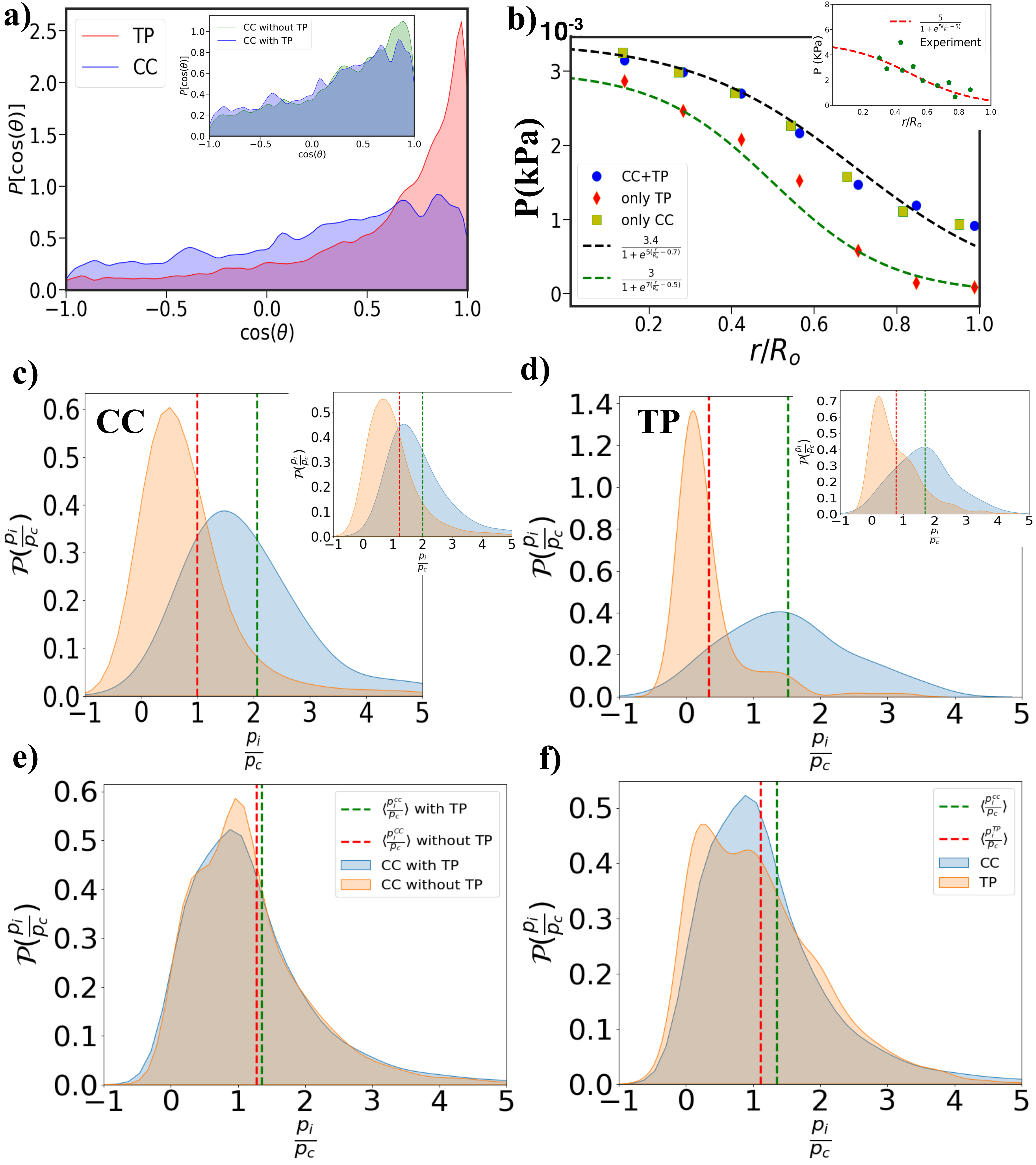}} 
\vspace{-.1 in}
\caption{ {\bf (a)} Distribution of $\cos(\theta)$ for the TPs and CCs.  $\theta$ is the angle between two consecutive steps along a  trajectory. The red (blue) plot is for the TPs (CCs). The distribution is skewed to $\cos(\theta)>0$, indicative of persistent motion. Extent of skewness is greater for TPs than CCs. The inset shows the $P[\cos (\theta)]$ of CCs with and without TPs. {\bf (b)}{\color{black}  Pressure as a function of the radial distance ($r$ scaled by $R_0$: tumor radius $\approx 100 \mu m$) from tumor center. The blue circles correspond to local pressure measured using both CCs and TPs. The red data points is local pressure measured using just the TPs using simulations containing both the CCs and TPs. The greenish yellow squares give the local pressure obtained in simulations  with only the CCs. The black and green dashed lines is logistic fit. The inset shows the radial pressure profile in experiments (green data points). The red dashed lines are logistic fits. Note that the magnitude of pressure measured in simulations and experiments are in qualitative agreement. (c) Pressure probability distribution, $\mathcal{P}(\frac{p_i}{p_c})$ for CCs. In the main figure, orange (blue) distributions correspond to stress distribution on CCs whose distance is greater (less) than 80 $\mu m$ (20 $\mu m$). Similarly in the inset,  orange (blue) distributions correspond to stress distribution on CCs whose distance is greater (less) than 60 $\mu m$ (30 $\mu m$). (d) Same as (c), except the distributions correspond to TPs. (e) The blue (orange) distribution corresponds to 
the stress distributions on the CCs in the presence (absence) of the TPs. (f) The blue (orange) distribution are the stress distributions on all the CCs (TPs) in the spheroid.} }
\label{theta_si}
\end{figure}

A mechanistic explanation for $\alpha_{TP}>\alpha_{CC}$ can be derived  from the simulations. To illustrate the difference between the dynamics of the TPs and CCs, we calculated the angle ($\theta$) between two consecutive time steps in a trajectory. We define $\cos(\theta(t,\delta t))=\frac{\delta {\bf r}(t+\delta t) \cdot \delta {\bf r}(t)}{|\delta {\bf r}(t+\delta t)||\delta {\bf r}(t))|}$, where $\delta {\bf r}(t)={\bf r}(t+\delta t)-{\bf r}(t)$. Figure(\ref{theta_si}a) shows the ensemble and time averaged $\cos \theta$ distribution for $\frac{\delta t}{\tau} = 1$. If the motion of the CCs and TPs were diffusive, the distribution of $\cos \theta$ would be uniformly distributed from -1 to 1. However, Figure(\ref{theta_si}a) shows that $P(\cos\theta)$ is skewed towards unity implying that the motion of both the TPs and CCs are persistent. Interestingly, the skewness is more pronounced for the TPs compared to the CCs (Figure(\ref{theta_si}a)). {\color{black} The enhanced persistence of TP motion due to SGAF  is also reflected in the FFA(t) displayed in Figure(\ref{gr_ffa}b). The slower decay of FFA(t) (Figure(\ref{gr_ffa}b)) in the TPs compared to the CCs is neglected in the distribution of $P[cos(\theta)]$, shown in Figure(\ref{theta_si}a).}  During each cell division, the motion of the CCs is randomized, and hence the persistence is small compared to TPs. This explains the hyper-diffusive (super-diffusive) for the TPs (CCs) and thus $\alpha_{TP}>\alpha_{CC}$ in the long time limit (see Figure(\ref{theta_si}a) which shows TPs movement is much more persistent than CCs). 
\bigskip

\noindent{\bf Monotonic decrease of pressure from the core to periphery of the MCS:~}{\color{black} To assess the effectiveness of the TPs as sensors of the local microenvironment, we calculated the radial pressure (Eq.(\ref{pressure1}) in the methods section) profile for both CCs and TPs at $t=7.5\tau$. In order to distinguish between the local pressure due to the TPs and CCs, we calculated pressure experienced  just by the TPs. Finally, we computed pressure in a system consisting of only the CCs.  By comparing the results from different simulations, we can unambiguously establish that TPs are excellent reporters of the local stresses within a single tumor.

From the results shown in  Figure(\ref{theta_si}b), we draw the following conclusions.
(i) All three curves  qualitatively capture the pressure profile found in the experiment \cite{Dolega17NC}. From the results in Figure(\ref{theta_si}b) several important conclusions that are relevant to experiments may be drawn. Pressure decreases roughly by a factor of four, as the distance, $r$,  from the center of the tumor increases. The pressure is almost constant in the core, with a decrease that can be fit using the logistic function, as the boundary of the tumor is reached  (Figure(\ref{theta_si}b)). The experimental data, shown in the inset in Figure(\ref{theta_si}b), was fit using a power law ($\sim r^{-\beta}$ with $\beta \approx 0.2$) (see Figure(4b) in the experiment \cite{Dolega17NC}). Both the fits account for the data with the important point being that the stress is non-uniform across the solid tumor. 

(ii) The experiments used reduction in volume of the polyacrylamide microbeads upon external compression ($=5KPa$) as local pressure sensors in colon carcinoma cells.  Although our simulations qualitatively reproduce the experimental pressure profile, the magnitude of the stress in the simulations is about three orders of magnitude less relative to experiments. The most likely reason is that in experiments high stress is uniformly applied by compressing the spherical tumor. In contrast, in our simulations pressure arises explicitly due to SGAF arising from an interplay of systematic forces, cell division, as well as mechanical feedback. The simulated MCS is not intended to mimic the characteristics of the colon carcinoma cells studied in the experiment \cite{Dolega17NC}. It should also be noted that the magnitude of the pressure is also dependent on the tissue type. For example, a recent experiment \cite{Mohagheghian18NC}, using fluorescently labeled water-soluble peptide-based microtubes, showed that the measured pressure $\approx 0.4 KPa$. The qualitative agreement between experiments and simulations (Figure(\ref{theta_si}b)) suggests that the dynamical rearrangement due to cell division and apoptosis and mechanical feedback controls the observed radial variations in the pressure profiles.  
 
 (iii) The higher value of pressure in the core of the MCS as the MCS grows is a consequence of jamming, which increase the local density. Because of the rapid cell division rate, the pressure experienced by cells in the interior is high (Figures \ref{theta_si}b, \ref{theta_si}c and \ref{theta_si}d). As a result, the local stresses in the core of the MCS do not have enough time to relax.The higher core density implies that pressure in the interior increases and exceeds $p_c$, which further inhibits proliferation, thus deceasing the number of cell divisions. The reduction in the number of cell divisions, which is a mechanism by which stress relaxation occurs, leads to increase in pressure at $r<R_0$ (Figures (\ref{theta_si}b, \ref{theta_si}c and \ref{theta_si}d)). As $r$ increases, the jamming effects diminish, local stress relaxation is faster resulting in the pressure on the cells being less than $p_c$. Consequently, the CCs proliferate, resulting in a decrease in self-generated stress, and consequently, a decreases in the magnitude of the pressure (Figures (\ref{theta_si}b, \ref{theta_si}c and \ref{theta_si}d)).
 
(iv) Most importantly, the CC pressure profiles are unaltered even in the presence of the TPs (compare blue circles and greenish-yellow squares in Figure \ref{theta_si}b, which shows that the latter does faithfully report the microenvironment of the CCs. In addition to the radial profile, we also show that the CC stress distribution is unaltered with and without TPs in Figure \ref{theta_si}e. The plot in Figure \ref{theta_si}f establishes that the stress distribution on the TPs roughly approximates the stress distribution on the CCs, implying that TPs can serve as reporters of not only the radial profile (Figure \ref{theta_si}b) but also the CC stress distribution. 

The simulations also provide plausible reasons for the decrease in  the pressure from the center to the boundary of the tumor. Pressure on the $i^{th}$ cell located at a distance ${ r}$ from the tumor center not only, depends on the total force exerted by the neighboring cells but also is controlled by cell division, which is a dynamic process. As the tumor evolves through cell division, the local density $\phi({\bf r})$ of the cells at small ${\bf r}$ increases. More importantly, cell division induces local stresses \cite {Doostmohammadi15SM}, which would relax on time scales that depend on $\phi({\bf r})$. Thus, cell division, jamming due to an increase in $\phi({\bf r})$, and stress relaxation times are intimately related. The smaller $\phi(\bf r)$ is faster in the stress relaxation time, and the smaller is the value of the pressure. It is clear that near the periphery of the tumor $\phi(\bf r)$ is smaller than at the core, which explains the origin of the pressure profile in Figure \ref{theta_si}b.
\bigskip

\noindent{\bf Conclusions:} We  used simulations and theory to elucidate the dynamics of inert tracer particles that are embedded in an evolving multicellular spheroid (MCS). Surprisingly, theory and simulations show that for $k_b t>1$ the TPs exhibit hyper-diffusive behavior, which does not depend on the details of interaction between the cells. In contrast, the exponent, $\alpha_{CC}$, characterizing the long-time behavior of $\Delta_{CC} (t)$ is nearly the same with or without the TP, and is also independent of the probe radius. This suggests that TPs may be used to probe the behavior of cancer cells within a solid tumor, which cannot be established unambiguously experiments alone. 

The most direct connections to experiments were made by showing that the simulated and measured stress profiles are qualitatively similar. This is unexpected because the protocol used in experiments (response to an isotropically applied stress to a spheroid) is completely different in the simulations in which stresses within the MCS are self-generated. The higher stress values in the interior of the spheroid imply that proliferation is suppressed in the core. The most important finding is that the CC pressure profiles are virtually identical independent of the  presence or absence if TPs (Figure(\ref{theta_si}(b)), which shows that the TPs are reporters of the CC microenvironment in a tumor.

An explanation for the radial dependence of local stresses in Figure(\ref{theta_si}(b) is related to the non-uniform proliferation of cells in the MCS. In the models, we used mechanical feedback to control the growth of the MCS.
This implies that if the local stresses exceed a critical value cells become dormant, and cannot growth until the neighboring cells rearrange to decrease the pressure on a labeled cell. However, as the MCS evolves,  the pressure on cells that are in the interior could exceed a critical value ($p_c$) due to jamming. In contrast, the pressure on cells near the MCS boundary is typically below $p_c$, which results in enhanced proliferation. It is the non-uniformity in the cell division rates due to mechanical feedback that not only is the cause of the radial variation in pressure in the MCS but also gives rise to the complicated radial dependent dynamics (glass-like slow dynamics in the interior to super diffusion near the periphery) \cite{sinha2019spatially}. 
The implication is that the effective diffusion constants of CCs as well as the TPs also vary with distance from the center of the MCS increases.\\

\noindent{\bf Methods:}\\
\bigskip

\noindent{\bf Numerical solution of the SIDs (Eq.(\ref{trdensity}) and Eq.(\ref{phi10})):} We numerically integrated Eq.(\ref{trdensity}) and Eq.(\ref{phi10}) by discretizing space in a cubic box with periodic boundary conditions.
The size of the 3D box with $L_x=L_y=L_z$ is varied from 300 to 500 $\mu m$ to ensure that the results do not depend on the system size. The 3D box was divided into grids, each with volume $\delta x \times \delta y \times \delta z$ with  $\delta x=\delta y= \delta z= 10 \mu m$. 
The parameters used in the numerical integration scheme are shown in Table I in the SI. Note that the precise shape of the box does not affect the numerical solutions nor does the value of $dx$ as long the solutions converge without numerical instability.

\medskip

\noindent{\bf Simulations:} {In order to test the theoretical predictions, and provide mechanistic insights underlying the unusual dynamics of the TPs in the MCS, we simulated a 3D tumor spheroid with embedded TPs. }{Following previous studies, we used an agent based model~\citep{Abdul17Nature,drasdo2005single} for the tumor spheroid. The cells  are treated as deformable spheres. The size of the CCs increase  with time, and divide into two identical cells upon reaching a critical mitotic radius ($R_m$). The mean cell cycle time, $\tau=54,000~ s$, which is a realistic value for fibroblast cells. The CCs can also undergo apoptosis with rate $k_a$. The birth rate, $k_b=\frac{1}{\tau}$ is large compared to the death rate ($\frac{k_b}{k_a}= 20$), which mimics the growth of the MCS. As in the theory, the TPs are  inert, and their sizes and the number are constant in the simulations. We use two different potentials for the CC-CC, CC-TP and TP-TP interactions.} \\


\noindent {\bf Hertz potential:} The form of the Hertz forces between the CCs is the same as in the previous studies \cite{Abdul17Nature,sinha2019spatially,drasdo2005single,schaller2005multicellular, pathmanathan2009computational}. The physical properties of the CC, such as the radius, elastic modulus, membrane receptor, and E-cadherin concentration characterize the strength of the inter-cellular interactions. The elastic forces between two spheres with radii $R_i$ and $R_j$, is given by,
\begin{equation}
F_{ij}^{el}=\frac{h_{ij}^{\frac{3}{2}}}{\frac{3}{4}(\frac{1-\nu_i^2}{E_i}+\frac{1-\nu_j^2}{E_j})(\sqrt{\frac{1}{R_i}+\frac{1}{R_j}})},
\label{hertzrepul}
\end{equation}
where $E_i$ and $\nu_i$ are, respectively, the elastic modulus and Poisson ratio of the $i^{th}$ cell. We assume that $E_i$ and $\nu_i$ are independent of $i$. Since the CCs or the TPs are deformable, the elastic force depends on the overlap, $h_{ij}$, between two cells.
The adhesive force, $F_{ij}^{ad}$, between the CCs is proportional to the area of contact ($A_{ij}$) \cite{palsson2000model}, and is calculated using, \cite{schaller2005multicellular},
\begin{equation}
F_{ij}^{ad}=A_{ij}f^{ad}\Lambda_0,
\label{hertzattra}
\end{equation}
where $\Lambda_0$ is  unity in the present study. 

Repulsive and adhesive forces in Eqs.(\ref{hertzrepul}) and (\ref{hertzattra}) act along the unit vector $\vec{n}_{ij}$ connecting  the centers of cells $j$ and $i$. Therefore, {the net force on} cell $i$ ($\vec{F}_{i}^{H}$) is given by the sum over the nearest neighbors [NN(i)],
\begin{equation}
\vec{F}_{i}^{H} = \Sigma_{j \epsilon NN(i)}(F_{ij}^{el}-F_{ij}^{ad})\vec{n}_{ij}.
\label{hertzian_force}
\end{equation}
To model the TP-TP and TP-CC interactions, we assume that the TPs are CC-like objects~\cite{Dolega17NC}. Therefore, CC-TP and TP-TP interactions are the same as CC-CC interactions.
The parameters used in the simulations using the Hertz potential (Eqs.(\ref{hertzrepul}) and (\ref{hertzattra})) are given in Table I in the SI. We initiated the simulations with 100 TPs and 100 CCs. The coordinates of the CCs and TPs were sampled using a normal distribution with zero mean, and standard deviation $50~\mu m$. The initial radii of the CCs and TPs were sampled from a normal distribution with mean $4.5~\mu m$, and a dispersion of  $0.5~\mu m$. \\

\noindent {\bf Gaussian potential:} In the theoretical treatment, we assumed that the CC-CC interaction is given by a sum of Gaussian terms (Eq.~(S2) in the SI). 
For this potential, the force ${\bf F}_{ij}^{G}$ on cell $i$, exerted by cell $j$, is,
\begin{equation}
{\bf F}_{ij}^{G}=\frac{1}{(2\pi)^{3/2}}[\frac{\nu e^{\frac{-r^2}{2\lambda^2}}}{\lambda^5}-\frac{\kappa e^{\frac{-r^2}{2\sigma^2}}}{\sigma^5}]{\bf r}
\label{gaussian_force}
\end{equation}
where ${\bf r}$ is ${\bf r}(i)-{\bf r}(j)$. 
We write $\lambda$ and $\sigma$ as $\lambda=\widetilde{\lambda}(R_i+R_j)$ and $\sigma=\widetilde{\sigma}(R_i+R_j)$, as the ranges of interactions corresponding to the repulsive and attractive interactions, respectively. {In our simulations, the CCs grow and divide, their radii change in time, and therefore $\lambda$ and $\sigma$ also change with time. However, since these interactions are short-ranged, we assume that they are constant in order to be consistent with the assumption in the theory. We fixed $\lambda=\widetilde{\lambda}(2R_d)$ and $\sigma=\widetilde{\sigma}(2R_d)$, where $R_d$ ($\approx 4 \mu m$) is the size of a daughter cell (introduced in the next section)}. For simplicity, we write the force as ${\bf F}_{ij}^{G}=[\frac{\widetilde{\nu} e^{\frac{-r^2}{2\lambda^2}}}{\lambda^2}-\frac{\widetilde{\kappa} e^{\frac{-r^2}{2\sigma^2}}}{\sigma^2}]{\bf r}$, where $\widetilde{\nu}=\frac{1}{(2\pi)^{3/2}}\frac{\nu}{\lambda^3}$ and $\widetilde{\kappa}=\frac{1}{(2\pi)^{3/2}}\frac{\kappa}{\sigma^3}$. 
 In the simulations, we fixed $\widetilde{\nu}=0.03$, {$\widetilde{\lambda}=0.28$}, $\widetilde{\kappa}=0.003$ and $\widetilde{\sigma}=0.4$.
 \medskip
 
\noindent{\bf Cell division, Dormancy and Apoptosis:} The CCs are either dormant or in the growth phase depending on the microenvironment, which is assessed using the value of the local pressure. The pressure on cell $i$ ($p_i$) due to $NN(i)$ neighboring cells is calculated using the Irving-Kirkwood equation, 
\begin{equation}
\label{pressure}
p_{i} =  \frac{1}{3V_i}\Sigma_{j \epsilon NN(i)}  {\bf F}_{ij}\cdot d{\bf r}_{ij},
\end{equation}
where ${\bf F}_{ij}$ is the force on the $i^{th}$ cell due to $j^{th}$ cell and $d{\bf r}_{ij} = {\bf r}_i- {\bf r}_j$. The volume of the $i^{th}$ cell ($V_i$) is $\frac{4}{3}\pi R_i^3$, where $R_i$ is the cell radius.   If $p_{i}$ exceeds a pre-assigned critical limit $p_c$ ($= 1.7$ Pa) the CC enters a dormant phase. The dormancy criterion serves as a source of mechanical feedback, which limits the growth of the tumor spheroid \cite{shraiman2005mechanical,alessandri2013cellular,conger1983growth,puliafito2012collective,gniewek2019biomechanical}. The volume of a growing cell increases at a constant rate, $r_V$. The cell radius is updated from a Gaussian distribution with the mean rate $\dot{R} = (4\pi R^2)^{-1} r_V$. Over the cell cycle time $\tau$, the cell radius increases as, 
\begin{equation}
r_V = \frac{2\pi (R_{m})^3}{3\tau},
\end{equation}
where $R_{m}$ is the mitotic radius. A cell divides once it grows to the fixed mitotic radius. To ensure volume conservation, upon cell division, we use $R_d = R_{m}2^{-1/3}$ as the radius of the daughter cells. The resulting daughter cells are placed at a center-to-center distance $d = 2R_{m}(1-2^{-1/3})$ (Figure(\ref{tracer_cell})). The direction of the new cell location is chosen randomly from a uniform distribution on a unit sphere. At times $\sim k_a^{-1}$, a randomly chosen cell undergoes apoptosis.  \\~\\
\medskip 
\noindent{\bf Equation of Motion:}  The equation of motion governing the dynamics of the TP and CCs is taken to be, 
\begin{equation}
\label{eqforce}
\dot{\vec{r}}_{i} = \frac{\vec{F}_{i}}{\gamma_i},
\end{equation}
where $\dot{\vec{r}}_{i} $ is the velocity of $i^{th}$ CC or TP, $\vec{F}_{i}$ is the force on $i^{th}$ CC/TP (see equation \ref{hertzian_force} and \ref{gaussian_force}), and $\gamma_i$ is the damping term.{ The details including the rationale for neglecting the random forces and inertial forces, are given elsewhere \cite{Abdul17Nature}. 

\medskip
\noindent{\bf Radial dependence of pressure:} We calculated the dependence of pressure $P(r)$ at a distance $r$ from the center of the tumor using,
\begin{equation}\label{pressure1}
P(r) = \frac{1}{n(r)} \sum_{i=1}^{n(r)} p_i,
\end{equation}
where $p_i$ is the pressure (Eq.(\ref{pressure})) on the $i^{th}$ cell located between $r$ and $r+dr$, and $n(r)$ is the number of cells in the same annulus. It is worth remembering that if the MCS is at equilibrium, $P(r)$ would be independent of $r$. Because in the simulations, cell proliferation is enhanced compared to the core of the MCS, there is more rapid relaxation of the stress at the MCS boundary, which result in reduced pressure.

In order to show that cell division is more likely to occur away from the core than in the interior of the MCS, we also calculated distribution ($P(p_i(r))$) of pressure experienced by cells in the annulus between $r$ and $r+dr$. We calculated $p_i(r)$ using the equation similar to Eq.(\ref{pressure1}) except in the $n(r)$ is the number of cells in the spherical volume between $r$ and $r+dr$.

\bigskip

\noindent\textbf{Acknowledgements:}
 \noindent 
We thank Xin Li, Abdul N. Malmi-Kakkada, Mauro Mugnai, Hung Nguyen, Rytota Takaki and Davin Jeong for useful discussions and comments on the manuscript.  This work is supported by the National Science Foundation (PHY 17-08128), and the Collie-Welch Chair through the Welch Foundation (F-0019).

\bibliographystyle{achemso}
\bibliography{tracer.bib}
\end{document}


\title{Supplementary Information: Tracer particles sense local stresses in an evolving multicellular spheroid without affecting the dynamics of cancer cells}
\author{Himadri S. Samanta}\affiliation{Department of Chemistry, University of Texas at Austin, TX 78712}
\author{Sumit Sinha}\affiliation{Department of Physics, University of Texas at Austin, TX 78712}
\author{D. Thirumalai}\affiliation{Department of Chemistry, University of Texas at Austin, TX 78712}

\date{\today}

\maketitle
\clearpage

{\bf Summary:} The Supplementary Information (SI) contains a number of items that are used to support the results in the main text. We first provide the theoretical results based on the numerical solution of the exact coupled stochastic integro-differential (SID) equations, describing the evolution of the inert tracer particles (TP) embedded in a growing multi-cellular spheroid (MCS). Using the numerical solution for the coupled SID equations and scaling ansatz, we  predict the exponents characterizing the dynamics of the TPs in the intermediate ($t \le \tau$ with $\tau$ being the time in which the CCs divide) and $t \gg \tau$.
We then provide additional results from the simulations, which not only validate the theoretical predictions but also elucidate the mechanism that drives the unusual dynamics of TPs, driven by the forces generated by division and apoptosis of the CCs, without affecting the CC dynamics.

\begin{figure}
\subfloat[]{\includegraphics[width=0.55\linewidth] {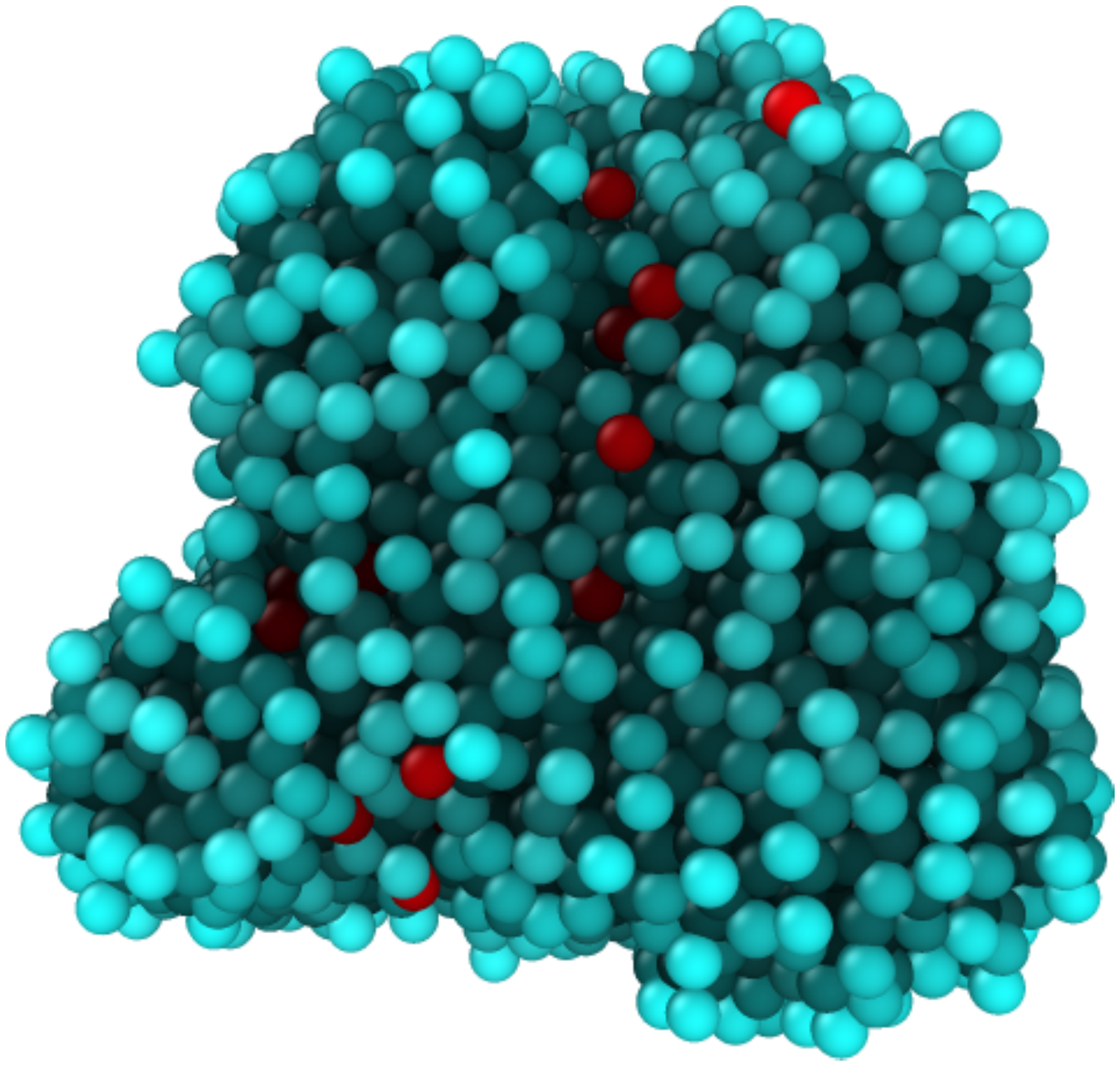}\label{tumor_pic1}}
\subfloat[]{\includegraphics[width=0.55\linewidth] {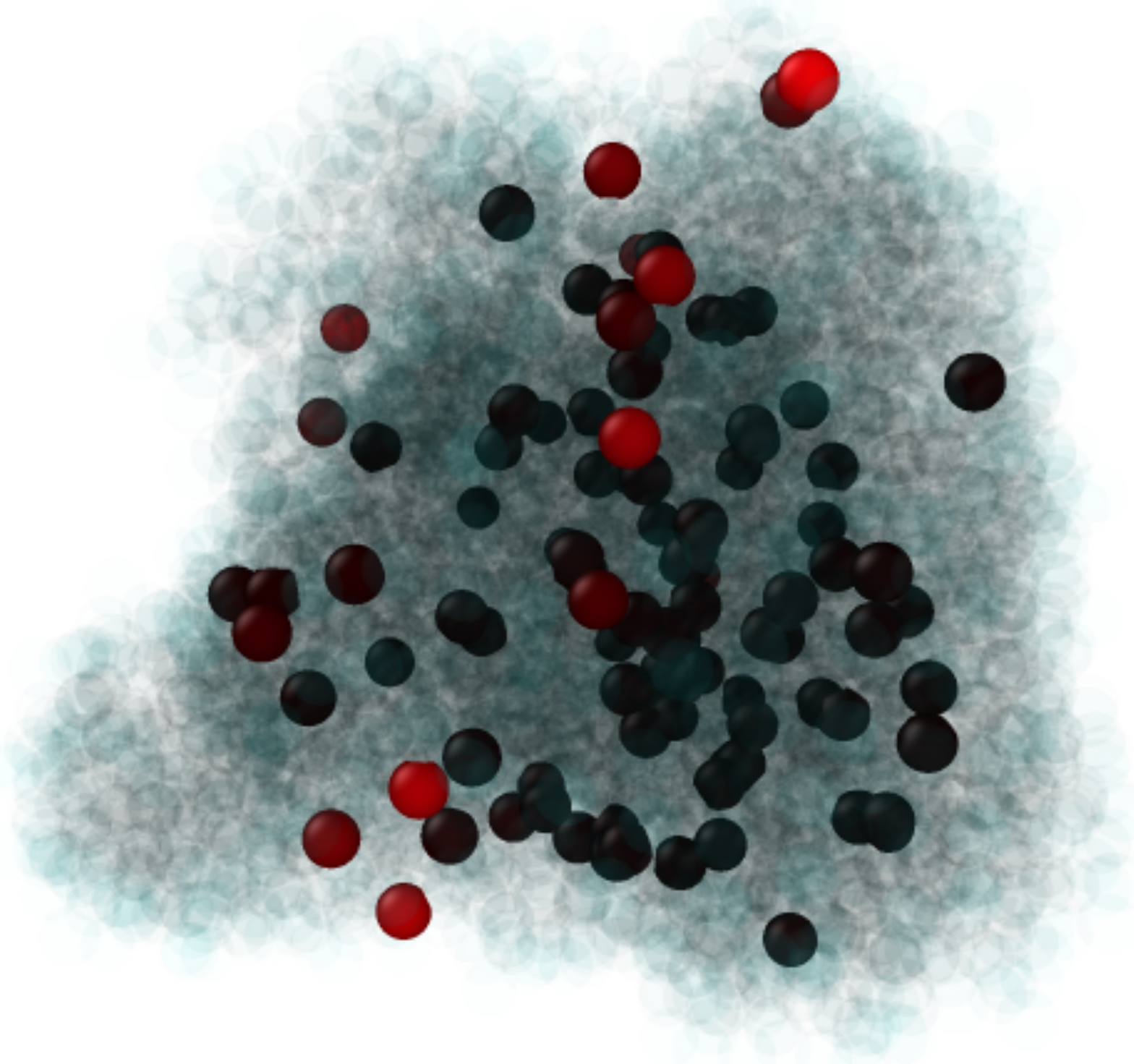}\label{tumor_pic2}} 
\caption{Snapshot from tumor simulations with embedded tracers. {\bf (a)} A 3D simulated spheroid consisting of approximately 4800 CCs and 100 TPs. The CCs are in cyan, and the tracers are in red.  {\bf (b)} The spheroid was rendered by making the CC cells transparent (light colored cyan) in order to show the interior of the spheroid. The TPs are opaque. Some of the TPs appear black because it is a depiction of a 3D image. The purpose of displaying these snapshots is to visually show that the TPs are randomly distributed within the multicellular spheroid, implying that their migration is largely determined by the forces arising from the CCs.}
\label{tumor_pic}
\end{figure}

{\bf Time-dependent equations for TP and CC densities:}
Let us consider  tracer particles (TPs) that are embedded  in a growing tumor spheroid, as shown Fig. \ref{tumor_pic}a. We model the short-range inter-cell interactions as a sum of two Gaussians that account for repulsion (arising from elastic forces) and attraction (mediated by E-cadherin). The latter accounts for volume excluded from the neighboring TPs and the cancer cells (CCs). In addition, the TPs and CCs are also subject to a random force characterized by a Gaussian white noise spectrum.
 We assume that the dynamics of the system, consisting of the CCs and TPs ({Figure \ref{tumor_pic} for snapshots generated in simulations}) can be described by {the  over damped} Langevin equation,
\begin{equation}
\frac{ d{\bf r}_i}{ dt}=-\sum_{j=1}^N \nabla U (|{\bf r}_i - {\bf r}_j|) +{\boldsymbol \eta}_i (t),
\label{eqn_of_motion}
\end{equation}
where $ {\bf r}_i$ is the position of a CC or a TP, and ${\boldsymbol \eta}_i(t)$ is a Gaussian random force with white noise spectrum. {The form of $U(|{\bf r}_i-{\bf r}_j|)$ between a pair of particles (can be either TP-TP, TP-CC or CC-CC) is taken to be }, 
\begin{equation}\label{potential}
U(|{\bf r}(i)-{\bf r}(j)|)=\frac{\nu}{(2\pi\lambda^2)^{3/2}}e^{\frac{-|{\bf r}(i)-{\bf r}(j)|^2}{2\lambda^2}}  
 -\frac{\kappa}{(2\pi\sigma^2)^{3/2}}e^{\frac{-|{\bf r}(i)-{\bf r}(j)|^2}{2\sigma^2}},
\end{equation}
where, $\lambda$ and $\sigma$ are the ranges of the repulsive and attractive interactions, and $\nu$ and $\kappa$ are the interaction strengths. Thus, the interactions involving the mixture of CCs and TPs are identical.
The potential in Eq.(\ref{potential}) is one of the model used in the simulations. 

{When Eq.~(\ref{eqn_of_motion}) is used to describe the TP dynamics}, the potential $U_{TP}$ contains both the TP-TP and TP-CC interactions with the corresponding attractive {(repulsive) interaction ranges being $\sigma_1 (\lambda_1)$ and $\sigma_2 (\lambda_2)$,  respectively.}
The potential $U_{CC}$ for the CCs mimics cell-cell adhesion (second term in the above equation), and the excluded volume interactions, and the CC-TP interactions.  

\floatsetup[figure]{style=plain,subcapbesideposition=top}

A closed form exact equation for the CC density, {$\phi({\bf r},t)=\sum_i \phi_i({\bf r},t)$} ( $\phi_i({\bf r},t)=\delta[\bf r-{\bf r}_i(t)]$) can be obtained using the well-known and standard Dean's method~\cite{Dean96JPA}. The formally exact time evolution of $\phi({\bf r},t)$ given by,
\begin{widetext}
\begin{eqnarray}
\label{phi10}
\frac{\partial \phi({\bf r},t)}{\partial t}&=& D_\phi \nabla^2 \phi({\bf r},t)+ {\bf \nabla }\cdot \left(\phi({\bf r},t)\int d{\bf r'}  [\psi({\bf r'},t){\bf \nabla}U_{CC-TP}({\bf r-\bf{r'}})
\right. \\ \nonumber && \left.
+ \phi({\bf r'},t){\bf \nabla}U_{CC}({\bf r-\bf{r'}})]\right)
+ \frac{k_a}{2} \phi(\frac{2k_b}{k_a}-\phi)+{\bf \nabla} \cdot \left(\eta_\phi({\bf r},t) \phi^{1/2}({\bf r},t)\right)
+\sqrt{k_b \phi+\frac{k_a}{2} \phi^2} f_\phi \, ,
\end{eqnarray}
\end{widetext}
where $\eta_\phi$ satisfies $<\eta_\phi({\bf r},t)\eta_\phi({\bf r'},t')>=2D_{\phi}\delta({\bf r}-{\bf r}')\delta(t-t')$ and $f_\phi$ satisfies $<f_\phi({\bf r},t)f_\phi({\bf r'},t')>=\delta({\bf r}-{\bf r}')\delta(t-t')$.
The term $\propto \phi(\phi_0-\phi)$ accounts for cell division (rate $=k_b$) and apoptosis (rate $=k_a$), with $\phi_0=\frac{2k_b}{k_a}$\cite{Doering03PA,Gelimson15PRL}.~The coefficient of $f_\phi$, given by~$\sqrt{k_b \phi+\frac{k_a}{2}\phi^2} $, is the strength of the noise corresponding to number fluctuations of the CCs, and is a function of the CC density. 
Similarly,
the evolution of the density of TP, $\psi({\bf r},t)=\sum_i \psi_i({\bf r},t)=\sum_i\delta[\bf r-{\bf r}_i(t)]$, may be written as,
\begin{widetext}
 \vspace{-.2 in}
\begin{eqnarray}\label{trdensity}
\frac{\partial \psi({\bf r},t)}{\partial t}&=&D_\psi \nabla^2 \psi({\bf r},t)+ {\bf \nabla }\cdot (\psi({\bf r},t)\int_{\bf r'} [\psi({\bf r'},t){\bf \nabla}U_{TP}({\bf r-\bf{r'}})
\\  \nonumber && 
+ \phi({\bf r'},t){\bf \nabla}U_{TP-CC}({\bf r-\bf{r'}})])+{\bf \nabla} \cdot \left(\eta_\psi({\bf r},t) \psi^{1/2}({\bf r},t)\right).
\end{eqnarray}
\end{widetext}
 where $\eta_\psi$ satisfies $<\eta_\psi({\bf r},t)\eta_\psi({\bf r'},t')>=2D_{\psi}\delta({\bf r}-{\bf r}')\delta(t-t')$.

{\color{black} The equations for $\phi({\bf r},t)$ and $\psi({\bf r},t)$ constitute coupled non-linear stochastic integro-differential (SID) equations, which are notoriously difficult to solve analytically. Usually approximations, often without establishing their validity, are used to solve the Dean's equation for one component fields. Here, we solve for the first time to our knowledge, by direct numerical integration of the coupled SIDs involving the fields $\phi({\bf r},t)$ and $\psi({\bf r},t)$ (Eqs.(\ref{phi10}) and (\ref{trdensity})). Because the thrust of the theoretical calculations is on the motility of the TPs and how they affect the  CC dynamics, we first numerically calculated the correlation functions associated with the TP density field ($\psi({\bf r},t)$). From the decay of the correlation functions and based by the expected scaling behavior, we computed the dynamical exponents for TPs, which is used to characterize the mean-square displacement (MSD $\sim t^{2/z}$), with $z $ being the dynamic exponent. 

By numerically integrating Eqs. \ref{phi10} and \ref{trdensity}, we  calculated the density-density correlations for the TPs, $C_{\psi \psi}(t) = \int d^3 {\bf r} C_{\psi\psi}({\bf r}, t)$ where 
$C_{\psi\psi}({\bf r}, t) = <\psi({\bf r}, t) \psi({\bf r}, t)>-<\psi({\bf r}, t)><\psi({\bf r}, t)>$. 
Based on the earlier works \cite{Himadri06PRE, Himadri06PLA, Himadri18PRE} we expect that the decay of TP density correlation function $C_{\psi\psi}(t)$ should decays as $C_{\psi\psi}(t) \sim t^{1-\frac{2}{z}-\frac{d}{z}}$ ($d$ is the spatial dimension). By fitting the expected scaling behavior for $C_{\psi \psi}(t)$ to the numerical solution to the SIDs the value of $z$ can be extracted. The MSD exponents are given by $2/z$.

Using the numerical results we calculated $C_{\psi \psi}(t)$ shown in Fig.(2a) in the main text. We find that both at short ($t<\tau$) and long times ($t>\tau$) $C_{\psi \psi}(t)$ decays as a power law.  By fitting the short time decay to $t^{1-2/z-d/z}\approx t^{-3/7}$ (green dashed line in Fig.(2a) in the main text) we extract $z=7/2$. The corresponding MSD exponent $\beta_{TP}=2/z=4/7$, which implies that at $t/\tau$ less than unity the TP motion 
is sub-diffusive. A similarly, at $t/\tau>1$, we find that $z=7/8$, which leads to the MSD exponent $\alpha_{TP}=16/7$. Thus, the TP dynamics changes sub-diffusive (glass-like) behavior at short times to hyper-diffusive at long times. The crossover occurs when the cell proliferation becomes relevant. A summary of the values of the exponent is given in Table II.

{\bf Control simulation: }As stated in the main text, TPs, regardless of their nature would be useful as local stress sensor only if they do not significantly alter the dynamics or the local stresses of the CCs. To ascertain if this holds we performed a variety of simulations by changing the TP size, the cell division time, and turning off the interaction between the TPs. The results of these simulations are described below.

\begin{table}[ht]
\caption{Parameters (except last three) used in the numerical integration of Eqs. \ref{phi10} and \ref{trdensity}. The values of the additional parameters (last three) used in the simulations.} 
\centering 
\begin{tabular}{c c c c} 
\hline\hline 
Definition & Parameters & Value  \\ [0.5ex] 
\hline 
Cell diffusion constant & $D_\phi$ & $10^{-8} \mu m^2/s$  \\ 
Tracer diffusion constant & $D_\psi$ & $10^{-8} \mu m^2/s$&  \\
Cell division rate & $k_{b}$ & $1/\tau,~\tau=54,000 s$ &\\
Cell apoptosis rate &$k_{a}$ & $0.1 k_b$&  \\
Box size & $L$  & $500~\mu m$&\\
Repulsive interaction range & $\lambda$  & $10~\mu m$&\\
Attractive  interaction range &$\sigma$& $ 10~\mu m$&\\
Repulsive interaction strength & $\nu$ & $103  \mu N \cdot \mu m^4$ & \\ [1ex] 
Attractive interaction strength & $\kappa $ & $100  \mu N \cdot \mu m^4$& \\ [1ex] 
Integration time step & $\delta t$ & $0.01 \tau$&\\ [1ex] 
{\color{black} Poisson ratio} & {\color{black}$\nu_i$} & {\color{black}0.5} &\\ [1ex] 
{\color{black} Elastic modulus} &{\color{black} $E_i$} & {\color{black}$10^{-3} MPa$} &\\ [1ex] 
{\color{black} Adhesive coefficient } & {\color{black}$f^{ad}$} & {\color{black}$10^{-4} \mu N/\mu m^2$} &\\ [1ex] 
\hline 
\end{tabular}
\label{table:nonlin} 
\end{table}

\floatsetup[figure]{style=plain,subcapbesideposition=top}
\begin{figure}
{\includegraphics[width=0.7\linewidth] {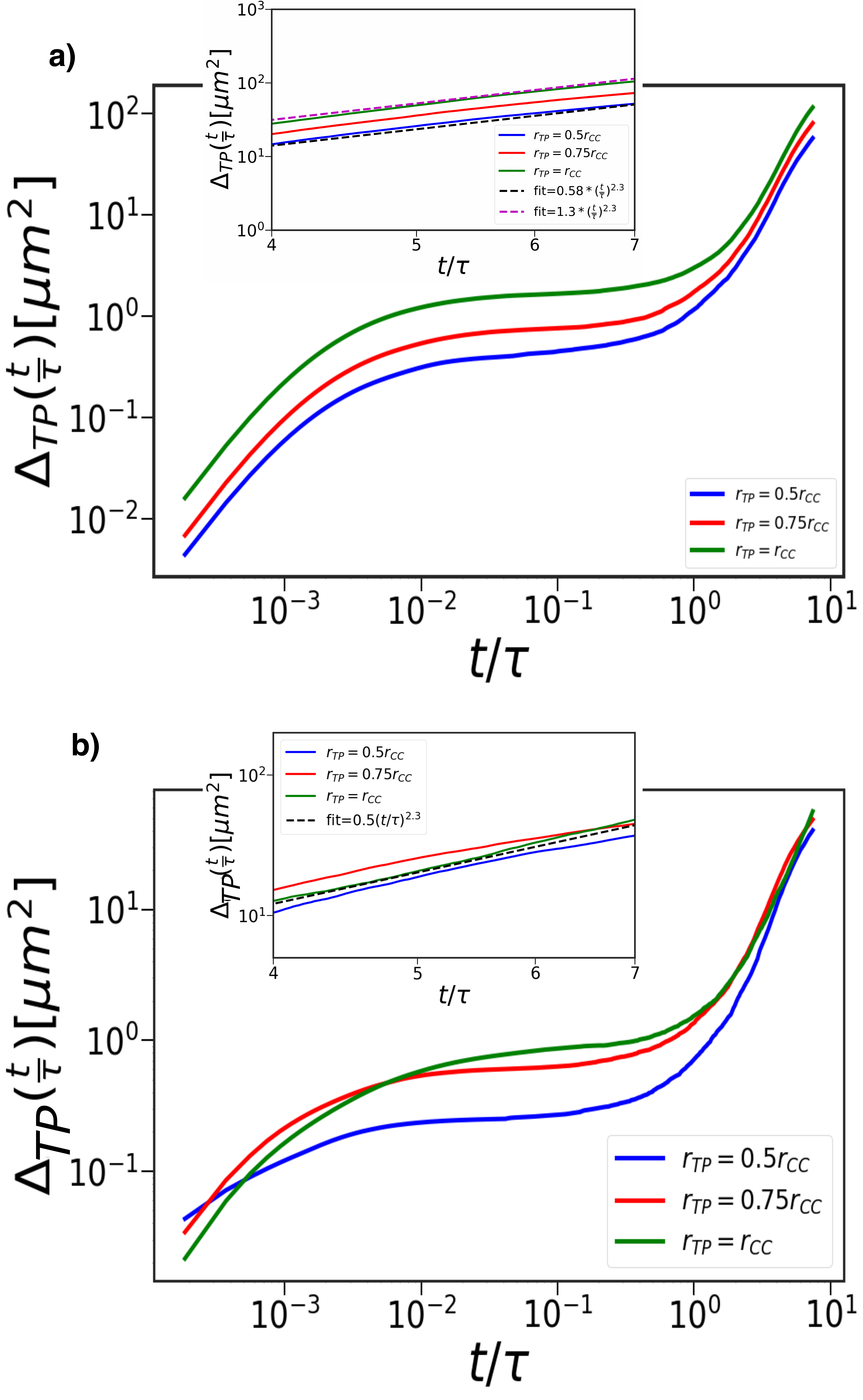}\label{fig3a}}
\caption{{\color{black} Dependence of the TP size on $\Delta_{TP}$ as a function of $t/\tau$. {\bf (a)} Data are for the Hertz potential (see equations (3)-(5) in the main text). From top to bottom, the curves correspond to decreasing TP radius ($r_{TP} = r_{CC}$ (green), $r_{TP} = 0.75r_{CC}$ (red) and $r_{TP} = 0.5r_{CC}$ (blue), where $r_{CC} =4.5 \mu m$ is average CC radius). TPs with larger radius have larger MSD values  in the intermediate time ($\frac{t}{\tau}\leq \mathcal{O}(1)$). In the inset, we focus on the hyper-diffusive regime. The black and magenta dashed line serves as a guide to the eye with  $\alpha_{TP}=2.3$  {\bf (b)} Same as (a) but the results are for the Gaussian interactions.}} .
\label{tracer_msd_diff_size}
\end{figure}

\floatsetup[figure]{style=plain,subcapbesideposition=top}
\begin{figure}
{\includegraphics[width=0.7\linewidth] {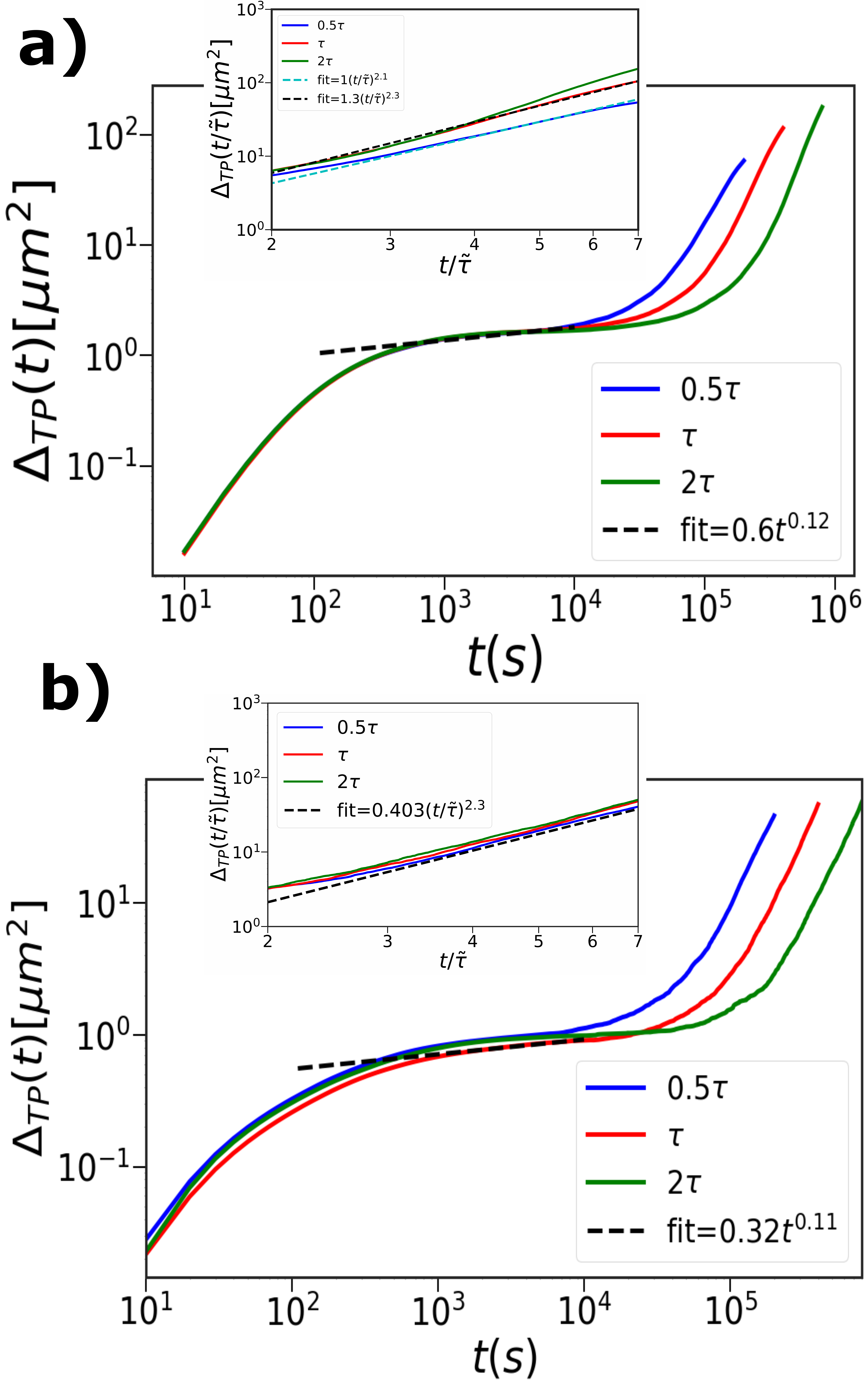}\label{fig3a}}
\caption{ {\color{black}Impact of cell division time $\tau$ on $\Delta_{TP}(t)$ as a function of time ($t$) for the two types of CC interactions. {\bf(a)} $\Delta_{TP}$, calculated using Eq.(5) in the main text, for the forces between the CCs and TPs. The curves are for 3 cell cycle times (blue -- $0.5\tau$, red --  $\tau$, and green -- $2\tau$). Time taken to reach the super-diffusive limit, which is preceded by a sub-diffusive (glass-like or jammed) regime, increases as $\tau$ increases. In the long-time, $\Delta_{TP}(t)$ exhibits hyper-diffusive motion ($\Delta_{TP}\sim t^{\alpha_{TP}}$  with $\alpha_{TP}>2$), which is shown in the inset for three $\tau$ values. The $x$-axis in the inset plot is scaled by $\frac{1}{\tilde{\tau}}$. {The values of $\tilde{\tau}$ are $0.5 \tau$, $\tau$, and $2 \tau$.} The black (cyan) dashed line shows exponent $\alpha_{TP}=2.3$ (2.1). The curve with 0.5 $\tau$ is best fit using $\alpha_{TP} =2.1$. {\bf(b)} Same as (a) except the Gaussian potential (Eq.(6) in the main text) is used in the simulations. Interestingly, $\alpha_{TP}$ does not change appreciably, implying that the long time dynamics is impervious to changes in the short range systematic interactions.}} .
\label{tracer_msd_birth}
\end{figure}

\floatsetup[figure]{style=plain,subcapbesideposition=top}
\begin{figure}
{\includegraphics[width=0.7\linewidth] {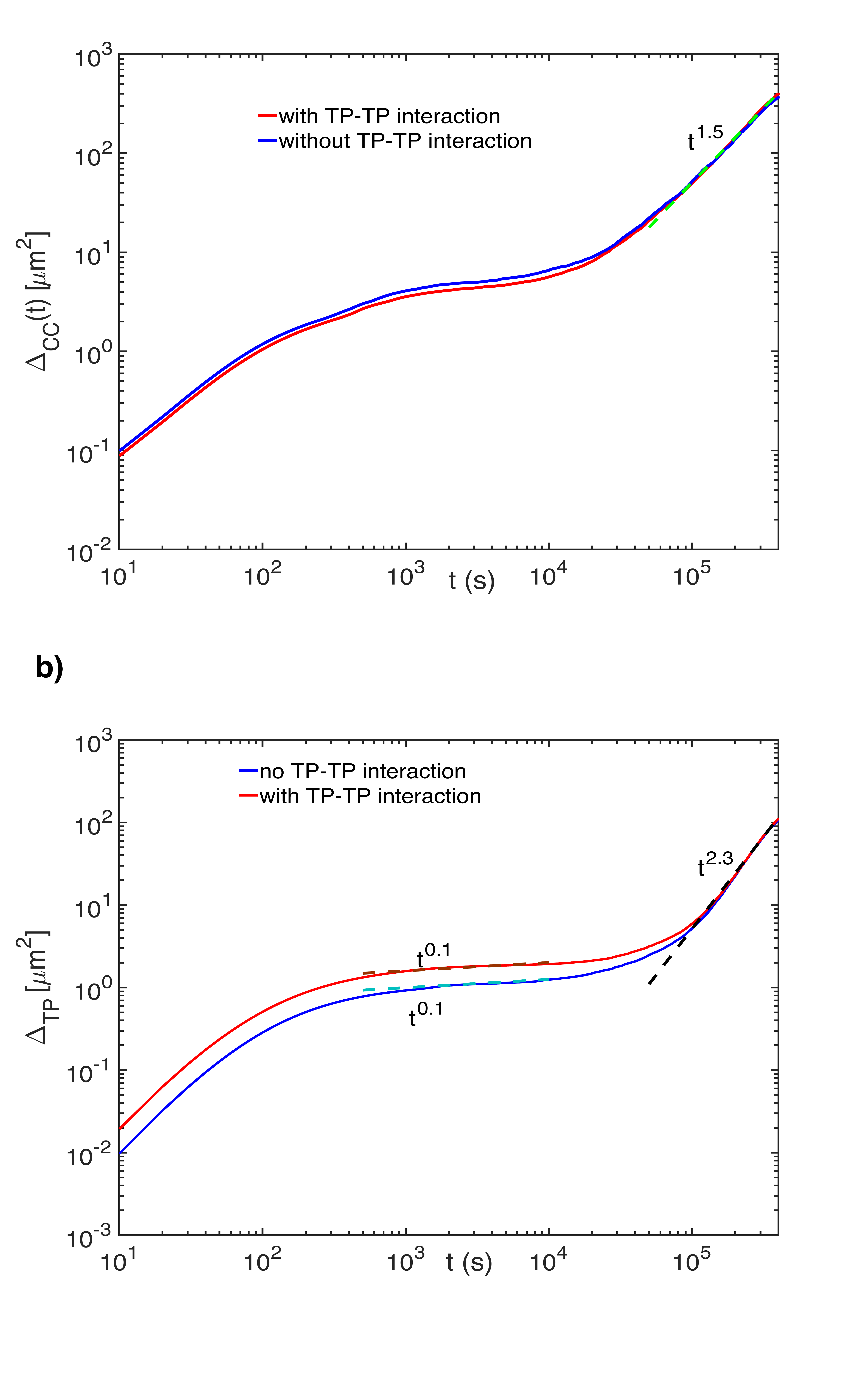}}
\caption{{\color{black} {\bf TP-TP interactions do not affect the long time dynamics of the TPs or CCs.} (a) $\Delta_{CC}$ with (red curve) and without (blue) TP-TP interactions. The TP-TP interactions play no role in the CC dynamics. The cyan dashed line shows $\alpha_{CC}=1.5$ for both the cases. (b)  $\Delta_{TP}$ with (red curve) and without (blue) TP-TP interaction. $\Delta_{TP}$ differs in magnitude at intermediate times. However, TP-TP interactions do not impact the long time hyper-diffusive dynamics of the TPs. }} .
\label{with_without}
\end{figure}

{{\bf $\alpha_{TP}$ is nearly independent of  the TP size:}} {We varied the radius of the TP ($r_{TP}$) from $0.5r_c$ to $r_c$, where $r_c=4.5 \mu m$ is the average CC radius.} Figure \ref{tracer_msd_diff_size} shows $\Delta_{TP}(t)$ as function of $t$ for the Hertz potential (Eq.(5) in the main text). Similar behavior is found for the Gaussian potential as well. In the intermediate time regime, larger sized TPs have higher MSD because they experience greater repulsive forces due to bigger excluded volume interactions. In the long time limit, $\Delta_{TP}$ exhibits hyper-diffusion (insets in Figure \ref{tracer_msd_diff_size}).{ The CC-TP interaction term,~${\bf \nabla }\cdot \left(\psi_1({\bf r},t)\int d{\bf r'} \phi_1({\bf r'},t){\bf \nabla}U({\bf r-\bf{r'}})\right)$, in Eq.(\ref{trdensity}) shows that the radius merely alters the interaction strength, and does not fundamentally change the scaling behavior. The conclusion that the values of $\alpha_{TP}$ do not change, anticipated on theoretical grounds, is supported by simulations.
 }

\floatsetup[figure]{style=plain,subcapbesideposition=top}
\begin{figure}
{\includegraphics[width=0.7\linewidth] {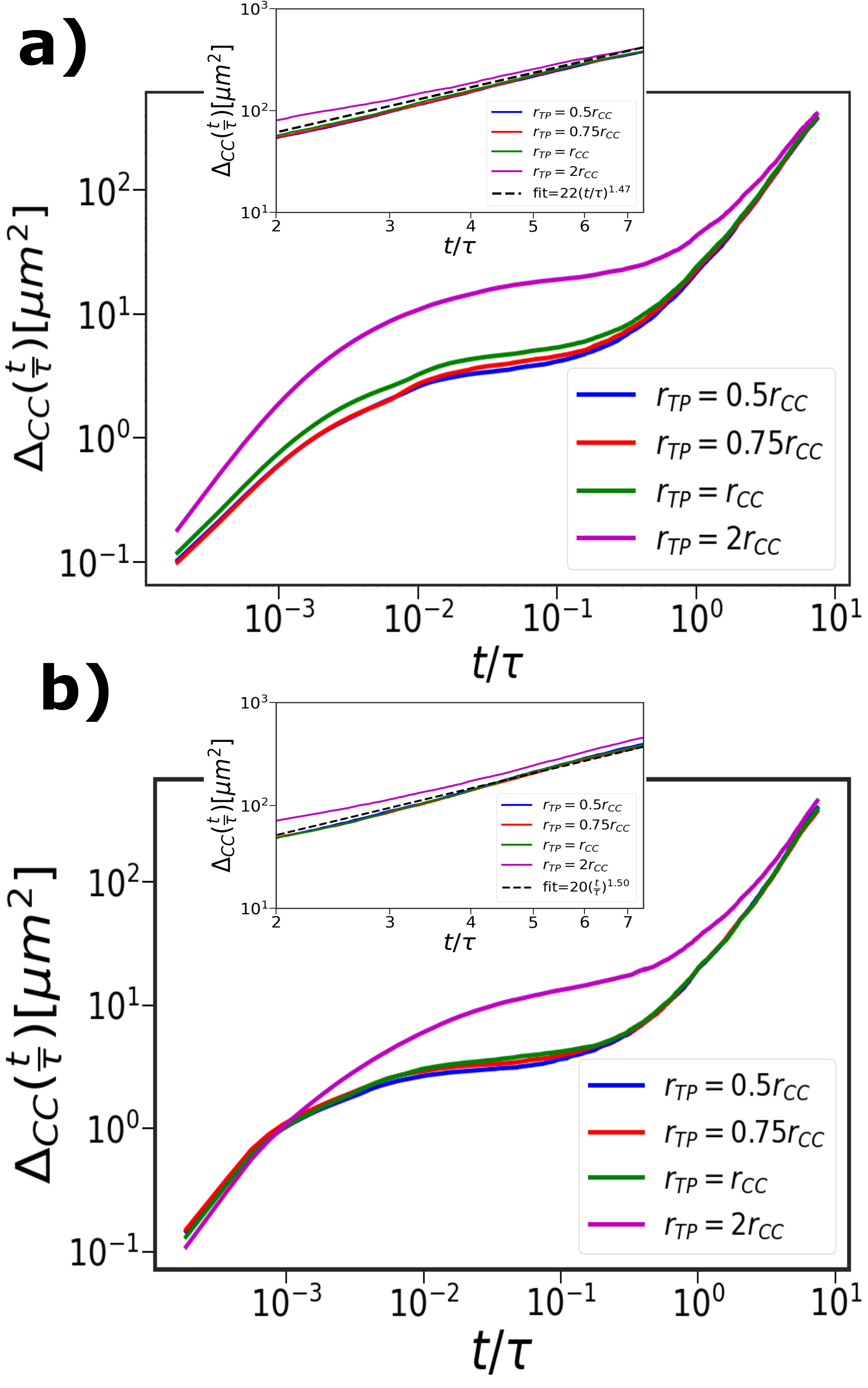}}
\caption{{\color{black} Influence of TPs on CC dynamics for two cell-cell potentials. (a) $\Delta_{CC}$, as function of $t/\tau$, using the Hertz potential (Eq.(5) in the main text). From top to bottom, the curves are for different values of the radius of the TPs (magenta $r_{TP} = 2 r_{CC}$, green $r_{TP} = r_{CC}$, red $r_{TP} = 0.75r_{CC}$  and blue $r_{TP} = 0.5r_{CC}$ (appears to be hidden), where $r_{CC} =4.5~\mu m$ is the average cell radius). In the intermediate times, $\Delta_{CC}(t)$ is larger for TPs with larger radius. The inset focuses on long times ($\frac{t}{\tau}>1$). The black line is a guide to show the value of  $\alpha_{CC}=$ 1.47. (b) Same as (a) but with the Gaussian potential. It is note worthy that the long time MSD exponent for the CCs is unaffected by the inert TPs, even after seven cell divisions.}} .
\label{fig4aa}
\end{figure}


{\bf CC cell division does not significantly alter the TP dynamics:} 
Fig.(\ref{tracer_msd_birth}) shows the influence of cell division on $\Delta_{TP}(t)$ for the Hertz and Gaussian potentials, described in the Methods section in the main text. It is clear that by varying cell division time from $(0.5-2)\tau$ the short time glassy dynamics as well as the hyper-diffusive behavior are maintained. Even the values of the exponents ($\beta_{TP}$ and $\alpha_{TP}$) are not significantly altered. 

{\bf Effect of TP-TP interactions:} Typically, in experiments the number of TPs in a given MCS is only about four or five. It is unlikely there is any interaction between the TPs. Even though we use 100 TPs they are sufficiently far away that $U_{TP-TP}\approx 0$. We performed simulations with and without TP-TP interactions. Fig.(\ref{with_without}) clearly shows that neither $\Delta_{TP}(t)$ nor $\Delta_{CC}(t)$ is affected by TP-TP interactions.  

{{\bf Influence of the TPs on the CC dynamics:}}
{\color{black} We calculated $\Delta_{CC}(t)$ as a function of $t$ using the Hertz (Gaussian) potential as a function of  the TP radius (Figure \ref{fig4aa}a (\ref{fig4aa}b)). For both the potentials, the values of $\Delta_{CC}(t)$ for $t< \tau$ increases as the TP sizes increase}.

{Before cell division, the number of CCs and TPs are similar, which explains the modest influence of the TPs on the dynamics of the CCs in the intermediate time regime}. The larger TPs experience stronger repulsion (the repulsive interaction is proportional to $R^2$) initially, which increases the magnitude of $\Delta_{CC}(t)$. The long-time dynamics is not significantly affected by the CC-TP interactions (see Figure \ref{fig4aa}). In the absence of the TPs, the CCs exhibit super-diffusion where the MSD scales as $t^{\alpha_{CC}}$ with $\alpha_{CC} =1.33$\cite{Abdul17Nature}.~In the presence of the TPs, the CC dynamics remains super-diffusive with $\alpha_{CC} \approx 1.45$, which shows that the TPs do not alter the CC dynamics significantly. The results in figures (\ref{tracer_msd_diff_size})-(\ref{fig4aa}) show that the influence of TPs on the CCs is minimal, which fully justifies their use as pressure sensors. The predictions for the CC dynamics made here remains to be tested. 

 \begin{table}
\begin{tabular}{ c|c|c|c} 
  & Theory & Hertz & Gaussian \\ 
 \hline
 $\beta_{TP}$ & 0.57 & 0.12 & 0.11 \\ 
 \hline
 $\alpha_{TP}$ & 2.28 & 2.30  & 2.30\\ 
 \hline
 $\alpha_{CC}$ & 1.45 & 1.47  & 1.50\\ 
 \hline
\end{tabular}
\caption{MSD exponents from theory and simulations.}
\end{table}

{{\bf Straigtness Index:}}~The Straightness Index (SI) is, given by, $SI_i=\frac{{\bf r}_i(t_f)-{\bf r}_i(0)}{\sum \delta {\bf r}_i(t)}$, where ${\bf r}_i(t_f)-{\bf r}_i(0)$ is the net displacement of $i^{th}$ TP or CC.  The denominator, $\sum \delta {\bf r}_i(t)$, is the total distance traversed. Fig. \ref{SI} shows that the TP trajectories are more straight (higher SI values on an average) or persistent over longer duration. During each cell division, the CCs are placed randomly causing the  trajectories to be less persistent, thus explaining the decreased persistence in their motion. In contrast, the TPs experience an impulsive kick in the vicinity of the CCs that undergo cell division. The kick results in a force on the TPs, which tends to produce persistent motion, thus giving rise to the higher SI values, as shown in Fig.(\ref{SI}).

\floatsetup[figure]{style=plain,subcapbesideposition=top}
\begin{figure}
{\includegraphics[width=0.8\linewidth] {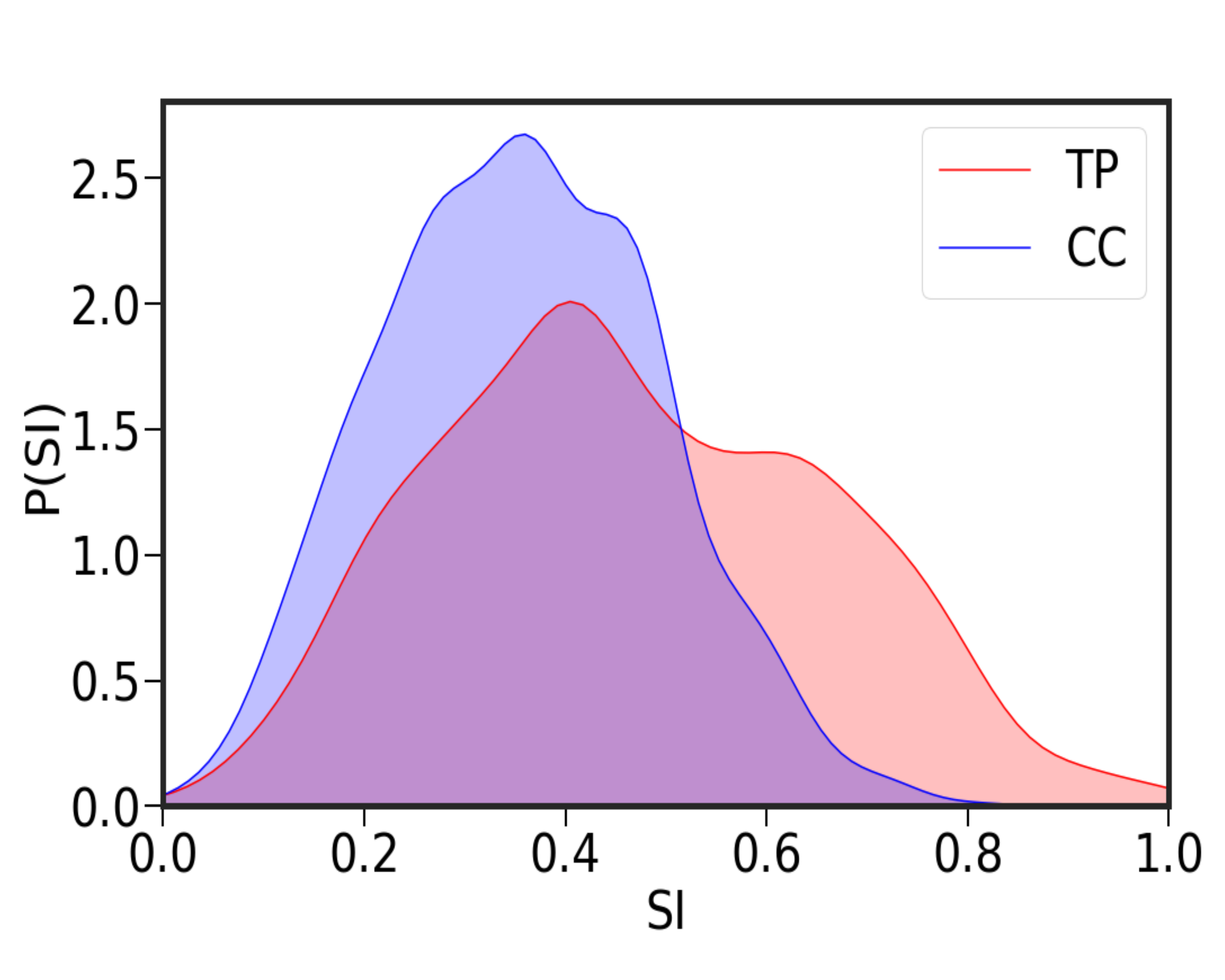}}
\caption{Distribution of the Straightness Index (SI). The red (blue) plot  for the TPs (CCs) shows that the TP trajectories are more rectilinear than the  CCs.}
\label{SI}
\end{figure}

\clearpage
\bibliography{tracer.bib}
\bibliographystyle{unsrt}